\documentclass[preprint,number]{elsarticle}
\usepackage[utf8]{inputenc}
\pdfoutput=1
\usepackage[a4paper]{geometry}
\usepackage{newunicodechar}
\newunicodechar{∂}{\partial}
\newunicodechar{∀}{\forall}
\newunicodechar{×}{\times}
\newunicodechar{α}{\alpha}
\newunicodechar{β}{\beta}
\newunicodechar{γ}{\gamma}
\newunicodechar{δ}{\delta}
\newunicodechar{ε}{\epsilon}
\newunicodechar{η}{\eta}
\newunicodechar{φ}{\varphi}
\newunicodechar{ϕ}{\phi}
\newunicodechar{λ}{\lambda}
\newunicodechar{ψ}{\psi}
\newunicodechar{ρ}{\rho}
\newunicodechar{ω}{\omega}
\newunicodechar{τ}{\tau}
\newunicodechar{θ}{\theta}
\newunicodechar{μ}{\mu}
\newunicodechar{ν}{\nu}
\newunicodechar{π}{\pi}
\newunicodechar{σ}{\sigma}
\newunicodechar{ξ}{\xi}
\newunicodechar{Γ}{\Gamma}
\newunicodechar{Ψ}{\Psi}
\newunicodechar{Ω}{\Omega}
\newunicodechar{Σ}{\Sigma}
\newunicodechar{Θ}{\Theta}
\newunicodechar{Λ}{\Lambda}
\usepackage{amsmath}
\usepackage{amsthm}
\usepackage{amssymb}
\usepackage{mathtools}
\usepackage{stmaryrd}
\usepackage{bbm}
\usepackage{tensor}
\usepackage{enumerate}
\usepackage{setspace}
\usepackage{verbatim}
\usepackage{ifpdf}

\ifpdf
\usepackage{hyperref}
\else
\fi

\DeclareFontFamily{OT1}{rsfs}{}
\DeclareFontShape{OT1}{rsfs}{m}{n}{ <-7> rsfs5 <7-10> rsfs7 <10-> rsfs10}{}
\DeclareMathAlphabet{\mycal}{OT1}{rsfs}{m}{n}
\newcommand{\mycalL}{{\mycal L}}
\newcommand{\mcL}{{\mycalL}}

{\catcode `\@=11 \global\let\AddToReset=\@addtoreset}
\AddToReset{equation}{section}

\newcommand{\R}{\ensuremath{\mathbb{R}}}

\newcommand{\1}{\ensuremath{\mathbbm{1}}}
\newcommand{\D}{\mathrm{d}}
\newcommand{\dx}{\mathrm{d}x}
\newcommand{\dy}{\mathrm{d}y}
\newcommand{\ds}{\mathrm{d}s}
\newcommand{\dt}{\mathrm{d}t}
\newcommand{\N}[1]{\mathcal{N}\indices{#1}}
\newcommand{\Ni}[1]{(\mathcal{N}^{-1})\indices{#1}}
\newcommand{\p}[1]{P\indices{#1}}

\newtheorem{thm}{Theorem}[section]
\newtheorem{cor}[thm]{Corollary}
\newtheorem{lemma}[thm]{Lemma}
\newtheorem{prop}[thm]{Proposition}
\theoremstyle{definition}
\newtheorem{definition}{Definition}

\usepackage{xspace}
\newcommand*{\eg}{e.g.\@\xspace}
\newcommand*{\ie}{i.e.\@\xspace}

\usepackage{color,soul}
\newcounter{mnotecount}[section]
\renewcommand{\themnotecount}{\thesection.\arabic{mnotecount}}
\newcommand{\mnote}[1]
{\protect{\stepcounter{mnotecount}}$^{\mbox{\footnotesize $%\!\!\!\!\!\!\,
\bullet$\themnotecount}}$ \marginpar{%\color{red}%
\raggedright\tiny\em
$\!\!\!\!\!\!\,\bullet$\themnotecount: #1} }
\newcommand{\ptcr}[1]{{\color{red}\mnote{{\color{red}{\bf ptc:}#1} }}}

\begin{document}
	\begin{frontmatter}
		\journal{Annals of Physics}
		\title{A New Class of Asymptotically Non-Chaotic Vacuum Singularities\tnoteref{t1,t2}}
		\tnotetext[t1]{This work is based on an MSc thesis under the supervision of Piotr T. Chru\'sciel.}
		\tnotetext[t2]{UWThPh-2015-17}
		\author[pk]{Paul Klinger}
		\ead{paul.klinger@univie.ac.at}
		\address[pk]{Faculty of Physics, Boltzmanngasse 5, 1090 Vienna, Austria}
		\begin{abstract}
			The BKL conjecture, stated in the 60s and early 70s by Belinski, Khalatnikov and Lifshitz, proposes a detailed description of the generic asymptotic dynamics of spacetimes as they approach a spacelike singularity. It predicts complicated chaotic behaviour in the generic case, but simpler non-chaotic one in cases with symmetry assumptions or certain kinds of matter fields.
			
			Here we construct a new class of four-dimensional vacuum spacetimes containing spacelike singularities which show non-chaotic behaviour. In contrast with previous constructions, no symmetry assumptions are made. Rather, the metric is decomposed in \emph{Iwasawa variables} and conditions on the asymptotic evolution of some of them are imposed. The constructed solutions contain five free functions of all space coordinates, two of which are constrained by inequalities.
			
			We investigate continuous and discrete isometries and compare the solutions to previous constructions. Finally, we give the asymptotic behaviour of the metric components and curvature.
		\end{abstract}
		\begin{keyword}
			General Relativity\sep Singularity\sep BKL\sep AVTD
		\end{keyword}
	\end{frontmatter}
	\section{Introduction}
	\subsection{Singularities in general relativity}
	When Albert Einstein presented his theory of General Relativity in 1915 he did not give any non-trivial exact solutions to its field equations. Due to the complicated non-linear structure of the equations, he did not expect any to exist and calculated physical predictions using perturbation theory \cite{Eisenstaedt1989}. To his surprise, less than a month later, Karl Schwarzschild sent him a letter containing the Schwarzschild metric, a spherically symmetric solution of the vacuum Einstein equations. It is given, in Schwarzschild coordinates, as
	\[
	\ds^2=-\left(1-\frac{2m}{r}\right)\dt^2+\left(1-\frac{2m}{r}\right)^{-1}\D r^2+r^2(\D θ^2+\sin^2 θ\, \D φ^2)\,.
	\]
	This solution contains, in these coordinates, an apparent singularity at $r=2m$ where the $rr$ component of the metric diverges. This is the event horizon, which Schwarzschild set as the origin of his coordinate system. The solution also contains a real singularity at $r=0$ which was found by David Hilbert in 1917. Hilbert considered both singularities real, as they could not be removed by an everywhere smooth and invertible coordinate transformation. In hindsight his requirement was too strict: The fact that in Schwarzschild coordinates the $rr$ component of the metric diverges at $r=2m$ simply means that these coordinates are badly chosen, indeed to transform from coordinates which don't show the apparent singularity to Schwarzschild coordinates requires a transformation which diverges at $r=2m$.
	
	In 1921 and 1922 Paul Painlevé and Allvar Gullstrand  independently discovered a spherically symmetric vacuum solution containing only a single singularity at $r=0$ \cite{Painleve1921,Gullstrand1922}. It was not realized  at the time that this solution can be obtained from the Schwarzschild one by a coordinate transformation, \ie it describes the same physical spacetime. This was finally discovered by Georges Lemaître in 1932, who also correctly identified the $r=2m$ singularity as an apparent singularity caused by the choice of coordinates \cite{Lemaitre1932}. The singularity at $r=0$ cannot be removed by a coordinate transformation as the Kretschmann scalar, given by $R^{αβγδ}R_{αβγδ}$, diverges there. This is a scalar quantity, constructed by contracting all indices of the Riemann tensor with itself, and is therefore independent of the chosen coordinate system.
	
	Despite this advance, the status of real singularities, such as the one appearing in the Schwarzschild or the cosmological FLRW solutions, was unclear. It was widely believed that they were an artifact of the symmetry assumptions made to obtain explicit solutions and had no relevance for the real world \cite{Penrose1969}. The idea was that, similarly to the Newtonian case, if matter was not perfectly symmetrically rushing towards a central point, the resulting angular momentum would prevent the formation of a singularity.
	
	The singularity theorems of Penrose and Hawking \cite{Penrose1965,Hawking1970} proved the opposite. They state that, given a trapped surface, an energy condition, and an assumption on the global structure of spacetime (e.g. no closed timelike curves), a singularity, in the sense of geodesic incompleteness, has to form. As small perturbations of an explicit solution containing a singularity would preserve the trapped surface, the perturbed solution also contains a singularity. These theorems, however, do not give any information about the nature of the predicted singularities, or about the behaviour of the metric near them. Indeed they do not even predict diverging curvature, only the existence of some geodesics, which cannot be extended beyond a finite value of the affine parameter along them.
	
	\subsection{The BKL conjecture}
	In a series of works, beginning in 1963, Belinski, Khalatnikov and Lifschitz (BKL) conjectured, based on heuristic arguments, that the dynamics of a generic spacetime containing a spacelike singularity would drastically simplify when the singularity is approached \cite{Lifschitz1963,Belinski1970}. They claimed that time derivatives of the metric would dominate compared to space derivatives, causing different spatial points to effectively decouple and turning the Einstein equations into a system of ODEs at each point. The solution of these ODEs is a generalisation of the Kasner metric, an explicit, homogeneous (but anisotropic) solution of the Einstein equations describing a spacetime which expands in some directions and contracts in others. It is given by
	\begin{equation}
	\ds^2=-\dt^2+\sum\limits_{j=1}^{d}t^{2p_j}\left(\dx^j\right)^2\,,
	\end{equation}
	with the constants $p_j$ fulfilling $\sum_j p_j=1$ and $\sum_j p_j^2=1$ (these conditions imply that at least one of the $p_j$ has to be negative, unless one is $1$ and all others $0$). The Kasner metric contains a singularity at $t=0$ as the Kretschmann scalar, given by
	\[
	K=R^{αβγδ}R_{αβγδ}=16\, t^{-4} \left(p_3{}^2-p_3{}^3\right)\,,
	\]
	diverges there.
	
	The behaviour predicted by BKL consists of a series of time periods (often referred to as Kasner epochs) during which the metric behaves at each spatial point as the Kasner metric, but with spatially varying exponents. At the end of a Kasner epoch the Kasner exponents $p_j$ change rapidly to a new configuration causing an ``oscillation'' as previously expanding directions contract. As the singularity is approached, the Kasner epochs get shorter and shorter and the transition between epochs becomes sharper.
	
	Chitre \cite{Chitre1972} and Misner \cite{Misner1994} introduced a representation of the BKL behaviour as a (chaotic) billiard motion in an auxiliary space of the same number of dimensions as the space part of the spacetime. A ``particle'', representing some parts of the metric, moves along straight, null, lines in a flat Lorentzian space and is elastically reflected off of (asymptotically) infinitely high potential walls. The straight line motion represents a Kasner epoch while the (asymptotically) sharp reflections correspond to the transitions between epochs. This billiard approach is described in detail by Damour, Henneaux and Nicolai in \cite{Damour2003}.
	
	Rigorous results concerning this chaotic case of the BKL conjecture are sparse: The only known example of a spacetime which shows the full chaotic BKL behaviour was constructed by Berger and Moncrief \cite{Berger:2000uf}. They applied a solution generating transformation to a homogeneous cosmological solution, yielding a $U(1)$ symmetric one. The resulting solution shows chaotic behaviour but it is very restricted, containing no free functions, and only three arbitrary constants.
	
	Numerical investigations do, however, provide strong evidence supporting the BKL conjecture \cite{Garfinkle2004}.  More recent simulations have shown that, while generically the spatial derivatives do become negligible, there are exceptional points at which they instead increase exponentially, giving spikes in the metric components \cite{Garfinkle2007}. In the class of Gowdy spacetimes explicit (non-chaotic) solutions exhibiting this behaviour have been found \cite{Rendall2001}. The appearance of these spikes hints at more complicated detailed behaviour within the general dynamics predicted by BKL.
	
	\subsection{Asymptotically simple behaviour}
	Belinski and Khalatnikov argued that coupling a massless scalar field to the Einstein equations would reduce the BKL behaviour to a simpler, non-oscillatory one, described by a single Kasner epoch, which is sometimes called AVTD (Asymptotically Velocity Term Dominated) \cite{Belinski1973}. This was rigorously proven, including the case of a stiff fluid, by Andersson and Rendall \cite{Andersson2001}.
	
	If a p-form field is added to the scalar one, the resulting behaviour is either simple (single Kasner epoch) or chaotic, depending on the coupling constant between them. This was shown by Damour, Henneaux, Rendall and Weaver \cite{Damour2002}.
	
	In the billiard picture, the addition of matter increases the dimension of the auxiliary space, as the particle describes not only the metric components but also the values of the matter fields. In addition, the evolution equations for the matter fields add additional potential walls. If a null line in the auxiliary space, which does not intersect any of the walls, exists, the resulting behaviour is simple, as a single Kasner epoch lasts up to the singularity.
	
	The addition of matter is not necessary for non-chaotic behaviour: Demaret, Henneaux and Spindel \cite{Demaret1985} argued, using similar heuristic arguments as BKL, that in 10 or more spatial dimension AVTD behaviour is generic.
	
	Even in lower dimensions, where BKL predict chaotic behaviour in the generic case, solutions which show non-chaotic behaviour exist. They are characterized by symmetry assumptions or conditions on their asymptotics. These assumptions cause some of the potential walls in the billiard picture to vanish at least asymptotically.
	
	This reduction was first proven for the polarized Gowdy subclass of the $T^2$ symmetric spacetimes by Chru\'sciel, Isenberg and Moncrief \cite{Isenberg:1989gq,Chrusciel:1990wn}. It was later extended to a larger class of Gowdy spacetimes by Kichenassamy and Rendall \cite{Kichenassamy1998} using a newly introduced ``Fuchsian'' method, which was then applied to more general $T^2$ symmetric spacetimes by Isenberg and Kichenassamy \cite{Isenberg1999}. An extension to so-called ``half-polarized'' $T^2$ spacetimes was achieved by Clausen and Isenberg \cite{Clausen2007}. Ames, Beyer, Isenberg and LeFloch \cite{Ames2013} extended the previous results on $T^2$ spacetimes to lower regularity. All results on $T^2$ symmetric spacetimes focused on the case of $3+1$ dimensions.
	
	$U(1)$ symmetric AVTD solutions, with only one Killing field, were constructed by Isenberg, Moncrief and Choquet-Bruhat in $3+1$ dimensions \cite{Isenberg2002,ChoquetBruhat:2005xs,ChoquetBruhat:2005rd}.
	
	The results obtained using Fuchsian methods do not necessarily provide generic solutions in the class of metrics considered, only the existence of families  of solutions which contain a number of arbitrary functions. As these functions specify the asymptotic behaviour of the solution, there is no obvious link with functions in the initial data. Within the class of Gowdy spacetimes, genericity of AVTD behaviour was proven by Ringström \cite{Ringstrom2008}.
	
	\subsection{This work}
	All previous results on simple behaviour in the vacuum case were obtained by starting with an ansatz for the metric which included one or more continuous symmetries. In the billiard picture this causes one or more of the walls to vanish identically at all times.
	
	Here a new class of non-chaotic vacuum solutions will be constructed without starting from such an ansatz. Instead, the decay of certain parts of the metric, defined by writing it in so-called \emph{Iwasawa variables}, will be required. This causes some of the walls in the billiard picture to vanish asymptotically. The approach is based on work by Damour and de Buyl \cite{Damour2008} who gave a precise statement of the BKL conjecture using this decomposition of the metric. Their work is an extension of \cite{Damour2003} by Damour, Henneaux and Nicolai. The new class of solutions includes the polarized Gowdy ones, but not the other classes mentioned above. It is at the same time more general, as it includes free functions which depend on all space coordinates, and more specific, as some asymptotically free functions in \eg the ``half-polarized'' $T^2$ case are here assumed to become constants in space.
	
	In sections \ref{sec:conventions} to \ref{sec:asym_evo} the relevant parts of \cite{Damour2008} are described: Section \ref{sec:conventions} describes the conventions and choice of gauge used and introduces the Iwasawa decomposition of the metric. In sections \ref{sec:actionham} and \ref{sec:walls} the action and Hamiltonian in the Iwasawa variable form are given, including the potential ``walls''. Section \ref{sec:fuchsthm} states the Fuchs theorem, which is the main tool used in the construction of the new class of solutions. In section \ref{sec:asym_evo} the evolution equations are written in Iwasawa form and the approach to constructing solutions with specified asymptotic behaviour, as used \eg by Rendall, is detailed. Finally, in section \ref{sec:new_class} the new class of solutions is constructed. In section \ref{sec:analysis} the new solutions are analysed: In sections \ref{sec:coord_free} and \ref{sec:killing} possible isometries of the solutions are investigated and in section \ref{sec:relprev} their relationship with previously known classes is described.
	
	In the appendices some of the calculations are given in more detail: In \ref{sec:ham_iwa_calc} the derivation of the Iwasawa form of the Hamiltonian is given in full. \ref{sec:equiv_eeq} contains comments on the form of the evolution equations used. In \ref{sec:calc_iwa_mom_const} the Iwasawa form of the momentum constraint equations is derived, following \cite{Damour2008}. In \ref{sec:const_evo} the evolution equations for the constraints are derived in the chosen gauge. \ref{sec:asym_beh} gives the asymptotic behaviour of the metric components and curvature for the new class of solutions. \ref{sec:cosm_const} shows that the new solutions can be constructed for arbitrary values of the cosmological constant.
		
	\section*{Acknowledgements}
	I would like to thank my supervisor, Piotr Chru\'sciel, for introducing me to this topic, for many stimulating discussions and for his continual support along the way.

	\section{Conventions, Iwasawa decomposition}\label{sec:conventions}
	We work with a $(-,+,\dots,+)$ signature. Greek indices $α,β,γ,\dots$ run from $0$ to $D=d+1$, Latin ones $a,b,c,\dots$ from $0$ to $d$. The metric (in $D=d+1$ dimensions) is written in the form
	\[
	\ds^2=-N(τ,x^i)^2 \D τ^2+g_{ij}(τ,x^i)\dx^i \dx^j\,,
	\]
	\ie with vanishing shift vector and lapse $N(τ,x^i)=\sqrt{\det g_{ij}}$. This \emph{pseudo-gaussian} gauge has the unusual property that changes of the spatial coordinates also change the slicing of the spacetime.
	
	The spatial metric is then decomposed into \emph{Iwasawa variables} $β^a$ and $\N{^a_i}$ as
	\[
	g_{ij}=\sum_{a=1}^d e^{-2β^a} \N{^a_i}\N{^a_j}\,.
	\]
	Here the $β^a$ and $\N{^a_i}$ are functions of all coordinates (including time) and $\N{^a_i}$ vanishes for all $a>i$ and is $1$ for $a=i$ (\ie $\N{^a_i}$ is upper triangular with ones on the diagonal). As the determinant of $\mathcal{N}$ is $1$, the determinant of the spatial metric only depends on the $β^a$ and is given by
	\begin{equation}
	\det g=e^{-2\sum_a β^a}\,.
	\end{equation} The $β^a$ are referred to as ``diagonal degrees of freedom'' while the $\N{^a_i}$ are the ``off-diagonal degrees of freedom'' (in fact both are relevant for all the metric components except $g_{11}$).
	
	This decomposition corresponds to a Gram-Schmidt orthogonalization of the coordinate coframe $dx^i$.
	
	The Iwasawa variables $β^a$ and $\N{^a_i}$ have the advantage that they explicitly separate parts of the metric which have different asymptotic behaviour: As we will see later, the $\N{^a_i}$ go to constants as $τ\to\infty$ while the $β^a$ approach linear functions.
	
	The Iwasawa coframe and its dual are defined as
	\begin{equation}\label{iwa_frame}
	θ^a=\N{^a_i}\dx^i \qquad\text{and}\qquad e_a=\Ni{^i_a}∂_i \,.
	\end{equation}
	The structure functions of the Iwasawa coframe, denoted $C\indices{^a_b_c}$, are defined by
	\[
	\Dθ^a=-\frac{1}{2}C\indices{^a_b_c}θ^b \wedge θ^c \quad\Leftrightarrow\quad [e_b,e_c]=C\indices{^a_b_c}e_a\,,
	\]
	and are given in terms of the $\N{^a_i}$ as
	\begin{equation}\label{structure_funcs}
	C\indices{^a_b_c}=\sum_{i,k}2\N{^a_k}\Ni{^i_{[b}}\Ni{^k_{c],i}}\,.
	\end{equation}
	In the $θ^a$ coframe the metric takes the diagonal form
	\[
	g_{ij}\dx^i \dx^j=\sum_{a=1}^d e^{-2 β^a} θ^a\!\otimes θ^a\,.
	\]
	
	\section{Action and Hamiltonian}\label{sec:actionham}
	Starting from the Einstein-Hilbert action
	\[
	S[\bar{g}_{μν}]=\int\D^Dx \underbrace{\sqrt{-\bar{g}}}_{\mathclap{=\sqrt{N^2\det g}=\det g}}\bar{R}\,,
	\]
	where $\bar{g}_{μν}$ is the spacetime metric with determinant $\bar{g}$ and $\bar{R}$ its Ricci scalar, the action can be written in Hamiltonian form as
	\[
	S[g_{ij}, π^{ij}]=\int\dx^0\int\D^dx\left(π^{ij}\dot{g}_{ij}-H\right)\,,
	\]
	where the $π^{ij}$ are the conjugate momenta to the spatial metric components, defined by
	\[
	π^{ij}=\frac{∂\mathcal{L}}{∂\dot{g}_{ij}}\,,
	\]
	and $H$ is the Hamiltonian density given by
	\begin{equation}\label{ham_gij}
	H=π^{ij}\dot{g}_{ij}-\mathcal{L}=π^{ij}π_{ij}-\frac{1}{d-1}π\indices{^i_i}π\indices{^j_j}-gR\,.
	\end{equation}
	This derivation is done e.g. in Appendix E of Wald \cite{Wald1984}.
	
	The Hamiltonian density can now be written in terms of the Iwasawa variables and their conjugate momenta $π_a$, corresponding to $β^a$, and $P\indices{^i_a}$, corresponding to $\N{^a_i}$ (note $P\indices{^i_a}=0$ for $a\geq i$) which are defined as
	\[
	π_a=\frac{∂\mathcal{L}}{∂\dot{β}^a}\quad\text{and}\quad P\indices{^i_a}=\frac{∂\mathcal{L}}{∂\dot{\mathcal{N}}\indices{^a_i}}\,.
	\]
	
	This gives
	\begin{equation}\label{ham_iwa}
	H=\overbrace{\frac{1}{4}G^{ab}π_a π_b}^{\mathcal{K}}+\overbrace{\sum_A c_A(\N{},P,∂_x β,∂^2_xβ,∂_x\N{},∂^2_x\N{})e^{-2ω_A(β)}}^{\mathcal{V}}\,,
	\end{equation}
	where $G^{ab}=(δ^{ab}(d-1)-1)/(d-1)$, $\N{}=(\N{^a_i})$ and $P=(\p{^i_a})$. The $d\times d$ matrix $G^{ab}$ is the inverse of $G_{ab}=-\sum_{c\neq d}δ^c_aδ^d_b$, which will appear later. In $d=3$ dimensions they are explicitly given by
	\[
	(G_{ab})=\begin{pmatrix*}[r]0&-1&-1\\-1&0&-1\\-1&-1&0\end{pmatrix*}\quad\text{and}\quad
	(G^{ab})=\frac{1}{2}\begin{pmatrix*}[r]1&-1&-1\\-1&1&-1\\-1&-1&1\end{pmatrix*}\,.
	\] The sum in the second term of (\ref{ham_iwa}) contains the potential ``walls'' which will be discussed in detail in the next section. The derivation of (\ref{ham_iwa}), including the individual terms in the second part, is given in \ref{sec:ham_iwa_calc}.
	
	The kinetic term $\mathcal{K}$ only contains the conjugate momenta of the diagonal $β^a$ variables, the ones for the $\N{^a_i}$ are included in the ``potential'' term $\mathcal{V}$. This makes sense because asymptotically the $\N{^a_i}$ tend to constants while the $β^a$ show linear behaviour, as will be demonstrated later.
	
	\section{The potential walls}\label{sec:walls}
	The structure of the potential term $\mathcal{V}$ in the Hamiltonian density \eqref{ham_iwa} is crucial for the asymptotic behaviour. It is that of a sum, with each term consisting of a prefactor which, importantly, does not depend on $β^a$ and an exponential term of the form $\exp(-2ω_A(β))$ where $ω_A$ is some linear form depending on the wall in question. Depending on the kind of wall, the index $A$ can be a single or a multi-index
	
	The walls are split into two categories: The so-called ``dominant'' and ``subdominant'' walls. The dominant ones are defined as the minimal set of walls such that if their linear forms are positive, all the others are as well. Crucially for the billiard picture, the coefficients $c_A$ are positive for the dominant walls.
	
	The form of \eqref{ham_iwa} allows the following ``billiard'' interpretation of the asymptotic dynamics (\eg \cite{Damour2003}): A ``particle'' with coordinates $β^a$ moves through a Lorentzian space with metric $G_{ab}$ (from the kinetic part $\mathcal{K}$ of the Hamiltonian) in a potential of the form $\mathcal{V}$. The behaviour of the summands in the potential is dominated by the exponential terms $\exp(-2ω_A(β))$. The $β^a$ can be decomposed as $β^a=ργ^a$ with $G_{ab}γ^aγ^b=-1$ and a heuristic argument, in analogy to the exact Kasner solution, gives $ρ\to\infty$ as $τ\to \infty$. In the limit, the potential walls become infinitely sharp as $-2ω_A(β)=-2ρω_A(γ)\to \pm \infty$. As long as $ω_A(β)>0$ the potential is negligible and the $β^a$ evolve linearly. At the points where $ω_A(β)$ becomes positive the potential diverges and, because $c_A>0$ for the dominant walls, the particle is reflected. The subdominant walls do not influence the behaviour as they lie behind the dominant ones.
	
	This picture depends on the assumption that $ρ\to\infty$ as $τ\to\infty$ and that none of the walls vanish (either completely or asymptotically). The following does not depend on these assumptions, as the billiard picture will not be used.
	
	In the vacuum case there are two types of potential walls: The ``symmetry walls'', coming from the kinetic terms of the off-diagonal metric components and the ``gravitational walls'' coming from the curvature term in the Hamiltonian density. The derivation of their exact form is given in \ref{sec:ham_iwa_calc}, here only the result is stated.
	
	\subsection{Symmetry walls}
	These come from the parts of the first two terms in the Hamiltonian density \eqref{ham_gij} which are not contained in the kinetic term $\mathcal{K}$ in (\ref{ham_iwa}). The part of $\mathcal{V}$ containing the symmetry walls is
	\begin{equation}\label{sym_walls}
	\sum_{a<b}\frac{1}{2}(\p{^j_a}\N{^b_j})^2 e^{-2(β^b-β^a)}\,,
	\end{equation}
	where the multi-index $A$ from (\ref{ham_iwa}) is $(a,b)$ and runs over all $a,b\in \{1,\dots,d\}$, $a<b$.
	The coefficients $(c_A)=(c_{ab})$ are given by $(\p{^j_a}\N{^b_j})^2/2$ and the linear forms $(ω_A)=(ω_{\text{sym }ab})$ by
	\begin{equation}
	ω_{\text{sym }ab}(β)=β^b-β^a \,.
	\label{sym_wall_forms}
	\end{equation} The walls with the forms $ω_{\text{sym } a\, a+1}$ are the dominant ones among the symmetry walls, because if they are positive then $β^{a+1}>β^a\,\, ∀a$ and therefore all the other $ω_{\text{sym }ab}(β)$, $a<b$ are positive as well.
	
	\subsection{Gravitational walls}
	These come from the curvature term in the Hamiltonian density (\ref{ham_gij}). The gravitational walls split into two classes:
	
	The contribution to $\mathcal{V}$ coming from the first class is given by
	\begin{equation}\label{dom_grav_walls}
	\sum_{a\neq b\neq c\neq a}\hspace{-0.9em}\frac{1}{4}(C\indices{^a_b_c})^2 e^{-2α_{abc}(β)}\qquad \text{with}\qquad α_{abc}=2β^a+\sum_{e\neq a,b,c}β^e\,,
	\end{equation}
	\ie the index $A=(a,b,c)$ is a multi-index running over all $a,b,c\in\{1,\dots,d\}$, $a\neq b\neq c\neq a$. For $d=3$ the sum in the expression for $α_{abc}(β)$ vanishes: There are only three possible values for the indices which are all occupied by $a$, $b$, and $c$, leaving no possible value for $e$. In this case only $α_{abc}(β)=2β^a$ remains.
	
	The second class of gravitational walls has a more complicated form, their contribution is given by
	\begin{equation}\label{subdom_grav_walls}
	-\sum_{a}F_a(∂_x^2β,∂_xβ,∂_xC,C) e^{-2μ_a(β)}\qquad \text{with}\qquad μ_a(β)=\sum_{c\neq a}β^c\,,
	\end{equation}
	and (all sums explicitly indicated)
	\begin{equation}\label{subdom_grav_coeff}\begin{split}
	F_a=&-2(β^a_{,a})^2-2β^a_{,a,a}\\
	&+\sum_{b}\bigg(-2(C\indices{^b_a_b})^2
	-4C\indices{^b_b_a}β^a_{,a}+4β^b_{,a}β^a_{,a}-(β^b_{,a})^2-2C\indices{^b_a_b}β^b_{,a}\\
	&\hspace{3.7em}+2β^b_{,a,a}+2C\indices{^b_a_b_{,a}}\\
	&\hspace{3.7em}+\sum_c\bigg(
	C\indices{^b_b_a}C\indices{^c_a_c}-β^b_{,a}β^c_{,a}-C\indices{^b_a_c}C\indices{^c_a_b}/2-2C\indices{^b_a_b}β^c_{,a}
	\bigg)\bigg)\,,
	\end{split}\end{equation}
	where the comma denotes the Iwasawa frame derivative $e_a$, defined in \eqref{iwa_frame} and given in terms of partial derivatives as $X_{,a}=\Ni{^i_a}∂_iX$.
	Here $A$ is a single index $a\in \{1,\dots,d\}$. This term contains second derivatives of $β^a$ and $\N{^a_i}$ (it contains first derivatives of $C\indices{^a_b_c}$ which contain first derivatives of $\N{^a_i}$) as expected from a curvature expression.
	
	The linear forms $μ_a$ of the second class can be written as a linear combination of the ones of the first class, $α_{abc}$, by
	\[
	μ_c=(α_{abc}+α_{bca})/2\,.
	\]
	This means the first class of walls is dominant and the second subdominant. This is fortunate as the coefficients of the second class of walls, $F_a$, can be negative while those of the first class, $(C\indices{^a_b_c})^2/4$, are always positive.
	
	\subsection{Complete Hamiltonian in Iwasawa variables}
	The complete Hamiltonian density in Iwasawa form (equation \eqref{ham_iwa} with the expressions for the walls inserted) is
	\begin{equation}\label{ham_iwa_full}\begin{split}
	H=&\frac{1}{4}G^{ab}π_a π_b+\sum_{a<b}\frac{1}{2}(\p{^j_a}\N{^b_j})^2 e^{-2(β^b-β^a)}+\hspace{-0.9em}\sum_{a\neq b\neq c\neq a}\hspace{-0.9em}\frac{1}{4}(C\indices{^a_b_c})^2 e^{-2(2β^a+\sum_{e\neq a,b,c}β^e)}\\
	&-\sum_{a}\bigg[
	-2(β^a_{,a})^2-2β^a_{,a,a}
	+\sum_{b}\bigg(-2(C\indices{^b_a_b})^2
	-4C\indices{^b_b_a}β^a_{,a}+4β^b_{,a}β^a_{,a}-(β^b_{,a})^2
	\\&\hspace{4.2em}-2C\indices{^b_a_b}β^b_{,a}
	+2β^b_{,a,a}+2C\indices{^b_a_b_{,a}}
	\\&\hspace{4.2em}+\sum_c\bigg(
	C\indices{^b_b_a}C\indices{^c_a_c}-β^b_{,a}β^c_{,a}-C\indices{^b_a_c}C\indices{^c_a_b}/2-2C\indices{^b_a_b}β^c_{,a}\bigg)\bigg)
	\bigg]e^{-2\sum_{c\neq a}β^c}\,,
	\end{split}\end{equation}
	with $C\indices{^a_b_c}=\sum_{i,k}2\N{^a_k}\Ni{^i_{[b}}\Ni{^k_{c],i}}$ and $G^{ab}=(δ^{ab}(d-1)-1)/(d-1)$ and where the derivative operator ``$_{,a}$'' is defined as $_{,a}=\Ni{^i_a}∂_i$.
	
	\section{Equations of motion and constraints}
	For a Hamiltonian density of the form $\mathcal{H}[q(x,t),p(x,t),∂_xq,∂^2_xq]$ the evolution equations are given by
	\begin{align*}\dot{q}(x,t)=&\frac{∂\mathcal{H}}{∂p}\,,\\
	\dot{p}(x,t)=&-\frac{∂\mathcal{H}}{∂q}+∂_m\frac{∂\mathcal{H}}{∂(∂_mq)}- ∂_m∂_n\frac{∂\mathcal{H}}{∂(∂_m∂_nq)}\,.
	\end{align*}
	
	Here the variation is taken after choosing lapse and shift, which depend on the metric (the lapse is given by $\sqrt{\det g}$). \ref{sec:equiv_eeq} shows that this does not change the resulting equations.
	
	In the case of the Iwasawa variable Hamiltonian (\ref{ham_iwa}) this leads to
	\begin{equation}\label{evo_eqs}
	\begin{aligned}
	∂_τβ^a=&\frac{1}{2}G^{ab}π_b\,,\\
	∂_τπ_a=&\sum_A \Bigg[2c_A(w_A)_ae^{-2w_A(β)}+∂_i\left(\frac{∂c_A}{∂(∂_i β^a)}e^{-2w_A(β)}\right)\\
	&\hspace{2.5em}-∂_i ∂_j \left(\frac{∂c_A}{∂(∂_i ∂_j β^a)}e^{-2w_A(β)}\right)\Bigg]\,,\\
	∂_τ\N{^a_i}=&\sum_A\frac{∂c_A}{∂P\indices{^i_a}}e^{-2w_A(β)}\,,\\
	∂_τP\indices{^i_a}=&\sum_A \Bigg[-\frac{∂c_A}{∂\N{^a_i}}e^{-2w_A(β)}+∂_j\left(\frac{∂c_A}{∂(∂_j\N{^a_i})}e^{-2w_A(β)}\right)\\
	&\hspace{2.5em}-∂_j∂_k \left(\frac{∂c_A}{∂(∂_j ∂_k\N{^a_i})}e^{-2w_A(β)}\right)\Bigg]\,,
	\end{aligned}
	\end{equation}
	where the components $(ω_A)_a$ of the linear form $ω_A$ appearing in the second equation are defined as $(ω_A)_a=∂ω_A(β)/∂β^a$.
	
	The Hamiltonian and momentum constraints are
	\begin{align}\label{full_ham_const}
	H&=0\,,\\
	\label{full_mom_const}\begin{split}
	-\frac{1}{2}H_a:\!&=\tilde{∂}_b\tilde{π}\indices{^b_a}+C\indices{^c_c_b}\tilde{π}\indices{^b_a}+C\indices{^d_a_c}\tilde{π}\indices{^c_d}-\frac{1}{2}(\tilde{∂}_a β^d)π_d\\
	&=0\,,\end{split}
	\end{align}
	where $\tilde{∂}_a=\Ni{^i_a}∂_i$ and
	\begin{equation}\label{pi_iwa_def}
	\tilde{π}\indices{^b_a}=
	\begin{cases}
	-\frac{1}{2}π_b\,,&\text{if } b=a\,,\\
	\frac{1}{2}\N{^b_i}P\indices{^i_a}\,,&\text{if } b>a\,,\\
	\frac{1}{2}e^{-2(β^a-β^b)}\N{^a_i}P\indices{^i_b}\,,&\text{if } b<a\,.
	\end{cases}\end{equation}
	The Iwasawa variable form (\ref{full_mom_const}) of the momentum constraints is derived in \ref{sec_full_mom_const_deriv}.
	
	\section{Fuchs theorem}\label{sec:fuchsthm}
	The following definition and theorem are by Choquet-Bruhat \cite[Appendix V, p. 636]{Choquet-Bruhat2008}, generalising the result of Kichenassamy and Rendall \cite{Kichenassamy1998}.
	
	\begin{definition}[Fuchsian System]\label{orig_fuchs_def}
		A system of partial differential first order equations on $V=M\times \R$, $M$ an analytic manifold which can be extended to a complex analytic manifold $\hat{M}$,
		\begin{equation}\label{orig_fuchs_form}
		t∂_tu+A(x)u=tf(t,x,u,D_xu)\,,
		\end{equation}
		with $f$ linear in the first order spatial covariant derivative $D_x u$, $A$ and $f$ extendable to holomorphic maps in $x$ and $u$ (on $\hat{M}$) and continuous in $t\in [0,T]$ is called Fuchsian if there exist $α<1$ and $Σ>0$ such that $σ^{A(z)}:=\exp (A(z)\log σ)$ satisfies
		\[
		\sup_{z\in \hat{M}}\left|σ^{A(t,z)}\right|σ^α\leq Σ\quad \text{for}\quad t\in [0,T]\,.
		\]
	\end{definition}
	\begin{lemma}
		A system of the form (\ref{orig_fuchs_form}), with $M$, $f$ as before, is Fuchsian if $A$ is uniformly bounded on $\hat{M}\times[0,T]$ with the real part of all its eigenvalues greater than $-1$.
	\end{lemma}
	\begin{thm}[Fuchs theorem]\label{thm:fuchs_theorem}
		A Fuchsian system has a unique solution $u$, analytic in $x\in M$, $C^1$ in $t$ and such that $u=0$ for $t=0$ in a neighbourhood of $\hat{M}\times \{0\}$.
	\end{thm}
	
	Replacing $t$ in \eqref{orig_fuchs_form} by $t'=t^{1/μ}$ gives
	\[
	μ^{-1}t'∂_{t'}u+A(x)u=t'^{μ}f(t(t'),x,u,D_xu)\,,
	\]
	and, as the eigenvalues of $μ^{-1}A$ are simply those of $A$ divided by $μ$, the following corollary holds.
	
	\begin{cor}
		The theorem holds for
		\[
		t∂_t u+A(x)u=t^μf(t,x,u,D_x u)
		\]
		if all eigenvalues $λ$ of $A$ fulfil $\text{Re}(λ)>-μ$.
	\end{cor}
	%\pk{New argument that $f$ can depend analytically on $D_x$. The dependence has to be analytic because $f$ has to be analytic in $u$ and $D_x$ is added as a new variable in $\hat{u}=(u,D_xu)$.}
	The condition that $f$ be linear in the spatial derivatives $D_x u$ can be relaxed to admit an arbitrary analytic dependence by adding $v:=D_x u$ as a new variable.
	Differentiating \eqref{orig_fuchs_form} gives an evolution equation for $v$,
	\begin{equation}
	t∂_t v+D_x A(x)u+A(x)v=tD_xf=t\left(\frac{∂f}{∂x}+\frac{∂f}{∂u}v+\frac{∂f}{∂v}D_x v\right)\,,
	\end{equation}
	which is linear in $D_x v$.
	Together with \eqref{orig_fuchs_form} this is a system of the form
	\begin{equation}\label{23III15.3}
	t∂_t \hat{u}+\hat{A}(x)=\hat{f}(t, x, \hat u, D_x \hat u)
	\end{equation}
	for $\hat{u}=(u,v)$ and with $\hat A$ the block lower triangular matrix
	\[
	\hat{A}=\begin{pmatrix}
	A&0\\
	D_xA&A
	\end{pmatrix}\,.
	\]
	The eigenvalues $λ$ of $\hat A$ fulfil
	\[
	0=\det(\hat A-\1 λ)=\det(A-\1 λ)^2
	\]
	and are therefore exactly the eigenvalues of $A$. Therefore the system \eqref{23III15.3} fulfils the conditions of definition \ref{orig_fuchs_def} and is Fuchsian, provided $\hat{f}$ depends analytically on $\hat u$, \ie $f$ depends analytically on $D_x u$, and $D_xA$ is uniformly bounded.
	\begin{cor}
		The existence theorem \ref{thm:fuchs_theorem} holds for $f$ depending analytically on $D_x u$ if $D_x A$ is uniformly bounded on $\hat{M}\times[0,T]$.
	\end{cor}
	
	A change of variables $t=e^{-μτ}\to τ$ in \eqref{orig_fuchs_form}, with $M$ a domain in $\R^n$, gives the form of the theorem used here.
	
	\begin{cor}\label{cor:fuchs_exp}
		A system of the form
		\begin{equation}\label{fuchs_form}
		∂_τu-A(x)u=e^{-μτ}\bar{f}(τ,x,u,D_xu)\,,
		\end{equation}
		with $A$ analytic in $x$ and uniformly bounded, $μ>0$, $\bar{f}$ analytic in $x$, $u$ and $D_xu$,
		continuous in $τ$ and bounded in $τ$ for $τ\to \infty$ and with all eigenvalues $λ$ of $A$ fulfilling $\text{Re}(λ)>-μ$ has a unique solution $u(x,τ)$ with $u(x,τ)\to 0$ as $τ\to \infty$.
	\end{cor}
	The matrices we will consider in the following will be constant and therefore the relevant conditions will be the boundedness of $\bar{f}$ as $τ\to \infty$ and the condition $λ>-μ$ on the eigenvalues of $A$.
	
	In order to obtain a more precise description of the decay of the solution, we define $\bar{u}=e^{ντ}u$, $0<ν<μ$. \eqref{fuchs_form} becomes
	\[
	∂_τ\bar{u}-(A+ν\1)\bar{u}=e^{-(μ-ν)τ}\bar{f}(τ,x,u(\bar{u}),∂_xu(\bar{u}))\,,
	\]
	which is again Fuchsian, as the conditions on $\bar{f}$ are unaffected and the eigenvalues of $A+ν\1$ are shifted up to compensate the change in $μ$. Therefore $\bar{u}=e^{ντ}u\xrightarrow{\scriptscriptstyle τ\to\infty}\;0$, \ie
	\begin{equation}\label{fuchs_decay}
	u=O(e^{-ντ})\,,\quad ∀\;\; 0<ν<μ\,.
	\end{equation}
	
	\section{Asymptotic evolution equations and differences} \label{sec:asym_evo}
	\subsection{Strategy and evolution equations}
	We will use the strategy introduced by Kichenassamy and Rendall in \cite{Kichenassamy1998} and used in \cite{Damour2008} to prove the existence of solutions of the Einstein equations with non-chaotic asymptotics. As a first step we consider a simplified system of evolution equations, which is supposed to model the asymptotic behaviour, and which can be easily solved. Then we write down the equations for the differences between solutions of this system and those of the full one, following from the full evolution equations. If this system can be shown to be Fuchsian (\ie if it is of the form (\ref{fuchs_form})) then, by the Fuchs theorem, a unique asymptotically vanishing solution exists. This implies that a solution of the full system of equations exists, which asymptotically approaches the specified solution of the simplified system.
	
	The constraints will be treated separately, in sections \ref{sec:asym_mom_const} and \ref{sec:asym_full_const}.
	
	Quantities relating to the asymptotic system will be marked with a subscript $_{[0]}$, \eg $β_{[0]}^a$.
	The Hamiltonian of the asymptotic system is obtained by discarding all wall terms in the full Hamiltonian (\ref{ham_iwa}), leaving only
	\begin{equation}\label{ham_asym}
	H_{[0]}=\frac{1}{4}G^{ab}π_{[0]a} π_{[0]b}\,.
	\end{equation}
	This gives the asymptotic evolution equations
	\begin{alignat*}{2}
	&∂_τβ^a_{[0]}=\frac{1}{2}G^{ab}π_{[0]b}\,,&\hspace{10em}& ∂_τπ_{[0]a}=0\,,\\
	&∂_τ\N{_{[0]}^a_i}=0\,,&\hspace{10em}& ∂_τP\indices{_{[0]}^i_a}=0\,,
	\end{alignat*}
	with solutions
	\begin{equation}\label{asym_sol}\begin{alignedat}{2}
	&β^a_{[0]}=p_\circ^aτ+β_\circ^a\,,&\hspace{10em}& π_{[0]a}=2G_{ab} p_\circ^b\,,\\
	&\N{_{[0]}^a_i}=\N{_\circ^a_i}\,,&\hspace{10em}& P\indices{_{[0]}^i_a}=P\indices{_\circ^i_a}\,.
	\end{alignedat}\end{equation}
	
	Now, the differences $\bar{β}^a$, $\bar{π}_a$, $\bar{\N{}}\indices{^a_i}$ and $\bar{P}\indices{^i_a}$ are defined as the real solutions minus the asymptotic ones (\eg, $\bar{β}^a=β^a-β_{[0]}^a$). Inserting them into the full evolution equations (\ref{evo_eqs}) gives the following equations for the differences:
	\begin{subequations}\label{diff_evo_eqs}\begin{align}
		∂_τ\bar{β}^a&-\frac{1}{2}G^{ab}\bar{π}_b=0\,,\label{diff_evo_b}\\
		\begin{split}
		∂_τ\bar{π}_a&=\sum_A \Bigg[2c_A(w_A)_ae^{-2w_A(β_{[0]})}e^{-2w_A(\bar{β})}+∂_i\left(\frac{∂c_A}{∂(∂_iβ^a)}e^{-2w_A(β_{[0]})}e^{-2w_A(\bar{β})}\right)\\
		&\hspace{3.5em}-∂_i∂_j \left(\frac{∂c_A}{∂(∂_i∂_jβ^a)}e^{-2w_A(β_{[0]})}e^{-2w_A(\bar{β})}\right)\Bigg]\,,\label{diff_evo_pi}\end{split}\\
		∂_τ\bar{\mathcal{N}}\indices{^a_i}&=\sum_A\frac{∂c_A}{∂P\indices{^i_a}}e^{-2w_A(β_{[0]})}e^{-2w_A(\bar{β})}\,,\label{diff_evo_n}\\
		\begin{split}
		∂_τ\bar{P}\indices{^i_a}&=\sum_A \Bigg[-\frac{∂c_A}{∂\N{^a_i}}e^{-2w_A(β_{[0]})}e^{-2w_A(\bar{β})}+∂_j\left(\frac{∂c_A}{∂(∂_j\N{^a_i})}e^{-2w_A(β_{[0]})}e^{-2w_A(\bar{β})}\right)\\
		&\hspace{3.5em}-∂_j∂_k \left(\frac{∂c_A}{∂(∂_j∂_k\N{^a_i})}e^{-2w_A(β_{[0]})}e^{-2w_A(\bar{β})}\right)\Bigg]\label{diff_evo_p}\,.\end{split}
		\end{align}\end{subequations}
	This system of equations is not directly in Fuchsian form, as the right-hand side contains second order spatial derivatives of the variables. By defining $B^a_j:=∂_j\bar{β}^a$ and $N\indices{^a_{ij}}:=∂_j\bar{\mathcal{N}}\indices{^a_i}$ these can be expressed as first order derivatives. This is only possible if the $\bar{β}$ and $\bar{\mathcal{N}}$ equations \eqref{diff_evo_b} and \eqref{diff_evo_n} do not contain spatial derivatives, as otherwise new second derivative terms would appear in the evolution equations for the new variables. The $\bar{β}$ equation \eqref{diff_evo_b} obviously doesn't contain spatial derivatives while for the $\bar{\mathcal{N}}$ equation \eqref{diff_evo_n} the sum over the walls only includes the symmetry walls with coefficients $(\p{^j_a}\N{^b_j})^2/2$, as the others are independent of $\bar{P}$.
	
	The additional evolution equations for the new variables are given by
	\begin{align}
	∂_τ B^a_j&=\frac{1}{2}G^{ab}∂_j\bar{π}_b\,,\label{diff_evo_B}\\
	∂_τ N\indices{^a_i_j}&=\sum_A ∂_j\left(\frac{∂c_A}{∂P\indices{^i_a}}e^{-2w_A(β_{[0]})}e^{-2w_A(\bar{β})}\right)\label{diff_evo_N}\,.
	\end{align}
	To ensure that the right-hand side of the equation for $B^a_j$ decays appropriately we replace $\bar{π}$ by $\tilde{π}$ defined as $\tilde{π}_a:=e^{ετ}\bar{π}_a$ with $ε>0$. The evolution equation for $B^a_j$ then becomes
	\begin{equation}
	∂_τ B^a_j=e^{-ετ}\frac{1}{2}G^{ab}∂_j\tilde{π}_b\,.\label{diff_evo_Bt}\\
	\end{equation}
	The evolution equation for $\tilde{π}_a$, which replaces equation \eqref{diff_evo_pi} is then
	\begin{equation}\label{diff_evo_pit}\begin{split}
	∂_τ\tilde{π}_a-ε\tilde{π}_a&=e^{ετ}\sum_A \Bigg[2c_A(w_A)_ae^{-2w_A(β_{[0]})}e^{-2w_A(\bar{β})}+∂_i\left(\frac{∂c_A}{∂(∂_iβ^a)}e^{-2w_A(β_{[0]})}e^{-2w_A(\bar{β})}\right)\\
	&\hspace{5em}-∂_i∂_j \left(\frac{∂c_A}{∂(∂_i∂_jβ^a)}e^{-2w_A(β_{[0]})}e^{-2w_A(\bar{β})}\right)\Bigg]\,.
	\end{split}\end{equation}
	with the additional term on the left-hand side and the exponential factor on the right-hand side coming from
	\[
	∂_τ\tilde{π}_a=∂_τ\left( e^{ετ}\bar{π}_a \right)=ε\tilde{π}_a+e^{ετ}∂_τ\bar{π}_a\,.
	\]
	
	The full system of equations is now given by the $2d+d(d-1)+d^2+d^2(d-1)/2$ equations \eqref{diff_evo_b}, \eqref{diff_evo_pit}, \eqref{diff_evo_n}, \eqref{diff_evo_p}, \eqref{diff_evo_N} and \eqref{diff_evo_Bt}.
	
	The asymptotic behaviour of the terms on the right-hand side is dominated by the exponential terms $\exp(-2ω_A(β_{[0]}))$. If $ω_A(β_{[0]})$ is strictly increasing with $τ$ for all $A$, \ie if $ω_A(p_\circ)>0$ for all $A$ (for all walls) the system fulfils the decay condition required by the Fuchs theorem (Corollary \ref{cor:fuchs_exp}). Equation \eqref{diff_evo_pit} includes the exponentially growing term $e^{ετ}$ but as $ε$ can be chosen arbitrarily small, and therefore smaller than the minimum of $ω_A(p_\circ)$, this does not affect the conditions.
	
	In order to be a Fuchsian system, the condition on the matrix $A$ also has to be fulfilled. For this system the matrix $A$ is given by
	\[\begin{pmatrix}
	0_d&(G^{ab})/2&0_{d, d_\star}&0_{d, d_\star}&0_{d,d^2}&0_{d,d_{\star\star}}\\
	0_d&ε\1_d&0_{d,d_\star}&0_{d,d_\star}&0_{d,d^2}&0_{d,d_{\star\star}}\\
	0_{d_\star, d}&0_{d_\star, d}&0_{d_\star}&0_{d_\star}&0_{d_\star,d^2}&0_{d_\star,d_{\star\star}}\\
	0_{d_\star, d}&0_{d_\star, d}&0_{d_\star}&0_{d_\star}&0_{d_\star,d^2}&0_{d_\star,d_{\star\star}}\\
	0_{d^2, d}&0_{d^2, d}&0_{d^2,d_\star}&0_{d^2,d_\star}&0_{d^2}&0_{d^2,d_{\star\star}}\\
	0_{d_{\star\star}, d}&0_{d_{\star\star}, d}&0_{d_{\star\star},d_\star}&0_{d_{\star\star},d_\star}&0_{d_{\star\star},d^2}&0_{d_{\star\star}}
	\end{pmatrix}\,,\]
	where $d_\star=d(d-1)/2$, $d_{\star\star}=dd_\star=d^2(d-1)/2$, $0_d$ is the $d\times d$ zero matrix, $0_{d,d_\star}$ is the $d\times d_\star$ zero matrix and $\1_d$ is the $d\times d$ identity matrix.
	The eigenvalues of this matrix are $0$ and $ε$. The condition therefore requires $\text{Re}(0)=0>-ω_A(p_\circ)$ and is fulfilled if $ω_A(p_\circ)>0\quad ∀ A$.
	
	In addition to the conditions $ω_A(p_\circ)>0$, the asymptotic Hamiltonian constraint, defined as
	\begin{equation}\label{asym_ham_const}
	H_\circ=G_{ab} p_\circ^a p_\circ^b=0\,,
	\end{equation}
	also constrains the values of the $p_\circ^a$.
	For vacuum in dimension $d<10$ the conditions $ω_A(p_\circ)>0\,\,∀A$ cannot be satisfied together with the asymptotic Hamiltonian constraint $H_\circ=0$ \cite{Demaret1985}. Therefore it is expected (\eg \cite{Damour2008}) that the generic solution in the vacuum case is chaotic.
	
	In $d=3$ dimensions it is easy to see why the conditions are not compatible: The linear forms of the dominant walls (these are the only relevant ones) are
	\begin{equation}\label{d3_walls}
	ω_{\text{sym } 21}(p_\circ)=p_\circ^2-p_\circ^1\,,\qquad ω_{\text{sym }32}(p_\circ)=p_\circ^3-p_\circ^2\,,\qquad α_{123}(p_\circ)=2p_\circ^1\,,
	\end{equation}
	(two symmetry walls from \eqref{sym_wall_forms} and one dominant gravitational wall from \eqref{dom_grav_walls}).
	
	The asymptotic Hamiltonian constraint is
	\begin{equation}\label{d3_ham_const}
	H_\circ=-p_\circ^1p_\circ^2-p_\circ^1p_\circ^3-p_\circ^2p_\circ^3=0\,.
	\end{equation}
	The condition that the three linear forms \eqref{d3_walls} are greater than $0$ implies $p_\circ^3>p_\circ^2>p_\circ^1>0$ and therefore $H_\circ<0$ which conflicts with the Hamiltonian constraint \eqref{d3_ham_const}.
	
	\subsection{Asymptotic momentum constraints}\label{sec:asym_mom_const}
	
	The asymptotic momentum constraints are obtained from the full momentum constraints (\ref{full_mom_const}) by splitting $\tilde{π}\indices{^b_a}$ (defined in (\ref{pi_iwa_def})) into a strictly upper triangular part $\tilde{π}\indices{^b_a_{[+]}}$, a strictly lower triangular part $\tilde{π}\indices{^b_a_{[-]}}$ and a diagonal part $\tilde{π}\indices{^b_b}=-π_b/2$, and discarding $\tilde{π}\indices{^b_a_{[+]}}$. This gives
	\begin{equation}\label{asym_mom_const}\begin{split}
	-\frac{1}{2}H_{a[0]}=&\tilde{∂}_b \tilde{π}\indices{_{[0]}^b_a_{[-]}}-\frac{1}{2}\tilde{∂}_aπ_{[0]a}+C\indices{_{[0]}^c_c_b}\tilde{π}\indices{_{[0]}^b_a_{[-]}}+C\indices{_{[0]}^d_a_c}\tilde{π}\indices{_{[0]}^c_d_{[-]}}\\
	&-\frac{1}{2}C\indices{_{[0]}^c_c_a}π_{[0]a}-\frac{1}{2}C\indices{_{[0]}^d_a_d}π_{[0]d}-\frac{1}{2}(\tilde{∂}_aβ\indices{_{[0]}^d})π_{[0]d}=0\,,
	\end{split}\end{equation}
	where $\tilde{∂}_a=(\N{_\circ^{-1}})\indices{^i_a}∂_i$ and $C\indices{_{[0]}^a_b_c}$ are the structure functions of the asymptotic Iwasawa coframe, defined as in (\ref{structure_funcs}) but with $\N{^a_i}$ replaced with $\N{_{[0]}^a_i}=\N{_\circ^a_i}$.
	
	The only time dependent term in (\ref{asym_mom_const}) is $-\tilde{∂}_aβ_{[0]}^dπ_{[0]d}/2=-τ(\tilde{∂}_ap_\circ^d)G_{dc} p_{\circ}^c$. This term vanishes if $G_{ab}p_{\circ}^a\,p_\circ^b=0$, \ie if the asymptotic Hamiltonian constraint is fulfilled, because the matrix $G_{ab}$ is symmetric and constant:
	\[
	(\tilde{∂}_ap_\circ^b)p_\circ^cG_{bc}=G_{bc}\frac{1}{2}\left((\tilde{∂}_ap_\circ^b)p_\circ^c+(\tilde{∂}_ap_\circ^c)p_\circ^b\right)=\frac{1}{2}(\tilde{∂}_a(\underbrace{p_\circ^b p_\circ^c G_{bc}}_{=p_\circ^bp_{\circ b}=0}))=0\,.
	\]
	The asymptotic constraints are therefore preserved under the asymptotic evolution given by (\ref{ham_asym}).
	
	\subsection{Relationship between asymptotic and full constraints }\label{sec:asym_full_const}
	
	We want to show that if the solution \eqref{asym_sol} of the asymptotic evolution system fulfils the asymptotic constraints, the corresponding solution of the full evolution equations fulfils the full constraints \eqref{full_ham_const}, \eqref{full_mom_const}.
	
	The evolution equations for the full constraints coming from the full evolution equations (\ref{evo_eqs}), in Iwasawa variables, are
	\begin{align}\label{ham_const_evo_eq}
	∂_τ H&=e^{-2\sum_b β^b}\sum_a\left(\tilde{∂}_a H^a-2\sum_{c} (\tilde{∂}_aβ^c)H^a\right)\,,\\
	\label{mom_const_evo_eq}
	∂_τ H_a&=\nabla_a H+\frac{H}{g}\tilde{∂}_ag\,,
	\end{align}
	with $H^a=e^{2β^a}H_a$ (derivation in \ref{sec:const_evo}). The right-hand side of (\ref{ham_const_evo_eq}) can be rewritten as
	\[
	\sum_a e^{-2μ_a(β)}\left[\tilde{∂}_aH_a+2(\tilde{∂}_aβ^a)H_a-2 \big(\tilde{∂}_a\sum_c β^c\big)H_a\right]\,,
	\]
	with $μ_a(β)=\sum_{b\neq a}β^b$ the subdominant gravitational wall forms.
	Defining $\bar{H}=e^{ητ}H$, with $η>0$, gives the system
	\begin{equation}\label{asym_full_fuchs}\begin{aligned}
	∂_τ\bar{H}-η\bar{H}&=\sum_a e^{η-2μ_a(β)}\left[\tilde{∂}_aH_a+2(\tilde{∂}_aβ^a)H_a-2 \big(\tilde{∂}_a\sum_c β^c\big)H_a\right]\,,\\
	∂_τH_a&=e^{-ητ}\bigg(\nabla_aH+\frac{H}{g}\tilde{∂}_ag\bigg)\,,
	\end{aligned}\end{equation}
	which is Fuchsian if $η<2μ_a(β)$. The term $\nabla_a H$ in the second equation is equal to $g\,\tilde{∂}_a(H/g)+H\, O(C\indices{^a_b_c})=g\,\tilde{∂}_a(H/g)+H\, O(1)$ where the first part comes from the density character of $H$ and the second from the connection coefficients in a non-coordinate basis. The system \eqref{asym_full_fuchs} is homogeneous and therefore the unique solution such that
	\begin{equation}
	\label{26XI14.1}
	\bar H\xrightarrow{\tau\to\infty} 0\quad\text{ and }  \quad  H_a\xrightarrow{\tau\to\infty}0
	\end{equation}
	guaranteed by the Fuchs theorem is $H_a=\bar{H}=H=0$. We therefore need to check that \eqref{26XI14.1} holds, \ie that the constraints are asymptotically fulfilled.
	
	%\ptcr{say something more about the exponential decay of the difference}
	The differences between the asymptotic constraints and the full ones consist only of terms which vanish asymptotically: The Asymptotic Hamiltonian (\ref{ham_asym}) is exactly the part $\mathcal{K}$ of the full Hamiltonian (\ref{ham_iwa}) which does not contain the exponential wall terms $\exp(-2ω_a(β))$, which go to zero if the conditions $ω_A(p_\circ)>0$ are fulfilled. The asymptotic momentum constraints were obtained from the full momentum constraints by discarding $\tilde{π}\indices{^b_a_{[+]}}=e^{-2(β^a-β^b)}\N{^a_i}P\indices{^i_b}/2$, $a>b$,  which is an exponentially decreasing term if the symmetry wall conditions (\ref{sym_wall_forms}) are fulfilled. This means that if the asymptotic constraints are fulfilled, the full constraints $H$ and $H_a$ vanish asymptotically. To verify \eqref{26XI14.1} we still need to make sure that the definition of $\bar{H}$ does not change the asymptotic behaviour. This is the case as $η$ can be chosen arbitrarily small, and therefore smaller than $2(β^a-β^b)$, $a>b$, while still preserving the Fuchsian form of \eqref{asym_full_fuchs}.
	
	This means, provided the solution of the asymptotic evolution equations satisfies the asymptotic constraints, \eqref{26XI14.1} is fulfilled. As the evolution equations for the full constraints are a homogeneous Fuchsian system, the unique solution which vanishes asymptotically is the zero solution. Therefore it suffices to impose the asymptotic constraints at one time (as they are preserved by the asymptotic evolution), to guarantee that the corresponding unique solution of the full evolution equations satisfies the full constraints at all times.
	
	\section{Construction of the new class of solutions}\label{sec:new_class}
	While generic solutions in the vacuum case are expected to be chaotic, there exist examples of vacuum spacetimes which show non-chaotic behaviour. As described in the introduction, all previous examples were at least $U(1)$ symmetric. These were constructed by starting with a symmetric ansatz for the metric, postulating asymptotic behaviour for its components and proving, via some sort of Fuchs theorem, that solutions with this asymptotic behaviour exist.
	
	Here, no symmetries of the metric will be assumed. The idea is to choose an ansatz for the $\N{^a_i}$ such that some of the walls in (\ref{diff_evo_eqs}) asymptotically vanish. This means that their linear forms can be negative but the resulting exponentially increasing term is countered by an exponential decrease of the coefficients $c_A$.
	
	\subsection{Ansatz and evolution equations}
	The following ansatz is chosen for $\N{^a_i}$:
	\begin{equation}\label{myansatz}
	\N{^a_i}(x^j,τ)=\N{_\circ^a_i}+e^{-γτ}\N{_s^a_i}(x^j,τ)\,.
	\end{equation}
	$\N{_\circ^a_i}$ is a constant, upper triangular matrix, with ones on the diagonal, which depends neither on space nor time. This ansatz for $\N{^a_i}$ will cause the dominant gravitational walls to vanish asymptotically. It can be simplified to $\N{_\circ}=\1$ by the space coordinate transformation $x^i\to y^i(x^j)$ defined by
	\begin{align*}
	y^1&=x^1+\N{_\circ^1_2}x^2+\N{_\circ^1_3}x^3\,,\\
	y^2&=x^2+\N{_\circ^2_3}x^3\,,\\
	y^3&=x^3\,,
	\end{align*}
	which does not affect the $β^a$.
	
	$P\indices{^i_a}$, $β^a$ and $π_a$ are decomposed as before in (\ref{asym_sol}), giving $d_\star+2d=d(d+3)/2$ functions of space $P\indices{_\circ^i_a}$, $p_\circ^a$ and $β_\circ^a$.
	
	As before, the second derivatives on the right-hand side of the evolution equations are eliminated by defining $B^a_j:=∂_j\bar{β}^a$ and $N\indices{_s^a_{ij}}:=∂_j\bar{\mathcal{N}}\indices{_s^a_i}$ and $\bar{π}$ is replaced by $\tilde{π}_a:=e^{ετ}\bar{π}_a$. The evolution equations for $\bar{β}^a$, $\tilde{π}_a$, $\N{_s^a_i}$, $\bar{P}\indices{^i_a}$, $B^a_j$ and $N\indices{_s^a_i_j}$ are now
	{\allowdisplaybreaks\begin{subequations}\label{my_diff_evo_eqs}
			\begin{align}
			\label{my_beta_eq}∂_τ\bar{β}^a-\frac{1}{2}G^{ab}\bar{π}_b&=0\,,\\
			\begin{split}
			\label{my_pi_eq}∂_τ\tilde{π}_a-ε\tilde{π}_a&=e^{ετ}\sum_A \Bigg[2c_A(w_A)_ae^{-2w_A(β_{[0]})}e^{-2w_A(\bar{β})}
			\\&\hspace{3.5em}+∂_i\left(\frac{∂c_A}{∂(∂_iβ^a)}
			e^{-2w_A(β_{[0]})}e^{-2w_A(\bar{β})}\right)
			\\&\hspace{3.5em}-∂_i∂_j \left(\frac{∂c_A}{∂(∂_i∂_jβ^a)}
			e^{-2w_A(β_{[0]})}e^{-2w_A(\bar{β})}\right)\Bigg]\,,
			\end{split}\\
			\label{my_n_eq}∂_τ\N{_s^a_i}-γ\N{_s^a_i}&=e^{γτ}\sum_A\frac{∂c_A}{∂P\indices{^i_a}}e^{-2w_A(β_{[0]})}e^{-2w_A(\bar{β})}\,,\\
			\begin{split}\label{my_p_eq}
			∂_τ\bar{P}\indices{^i_a}&=\sum_A \Bigg[-\frac{∂c_A}{∂\N{^a_i}}
			e^{-2w_A(β_{[0]})}e^{-2w_A(\bar{β})}
			\\&\hspace{3.5em}+∂_j\left(\frac{∂c_A}{∂(∂_j\N{^a_i})}e^{-2w_A(β_{[0]})}
			e^{-2w_A(\bar{β})}\right)
			\\&\hspace{3.5em}-∂_j∂_k \left(\frac{∂c_A}{∂(∂_j∂_k\N{^a_i})}
			e^{-2w_A(β_{[0]})}e^{-2w_A(\bar{β})}\right)\Bigg]\,,		\end{split}\\
			∂_τ N\indices{_s^a_i_j}&=e^{γτ}\sum_A ∂_j\left(\frac{∂c_A}{∂P\indices{^i_a}}e^{-2w_A(β_{[0]})}e^{-2w_A(\bar{β})}\right)\label{my_N_eq}\,,\\
			∂_τ B^a_j&=e^{-ετ}\frac{1}{2}G^{ab}∂_j\tilde{π}_b\,.\label{my_Bt_eq}
			\end{align}\end{subequations}}
	
	The additional term on the left-hand side of the $\N{_s^a_i}$ equation and the exponential factor on the right-hand side come from
	\[
	∂_τ\left( e^{-γτ}\N{_s^a_i} \right)=-γe^{-γτ}\N{_s^a_i}+e^{-γτ}∂_τ\N{_s^a_i}\,.
	\]
	
	The matrix $A$ is now
	\[\begin{pmatrix}
	0_d&(G^{ab})/2&0_{d, d_\star}&0_{d, d_\star}&0_{d,d^2}&0_{d,d_{\star\star}}\\
	0_d&ε\1_d&0_{d,d_\star}&0_{d,d_\star}&0_{d,d^2}&0_{d,d_{\star\star}}\\
	0_{d_\star, d}&0_{d_\star, d}&γ\1_{d_\star}&0_{d_\star}&0_{d_\star,d^2}&0_{d_\star,d_{\star\star}}\\
	0_{d_\star, d}&0_{d_\star, d}&0_{d_\star}&0_{d_\star}&0_{d_\star,d^2}&0_{d_\star,d_{\star\star}}\\
	0_{d^2, d}&0_{d^2, d}&0_{d^2,d_\star}&0_{d^2,d_\star}&0_{d^2}&0_{d^2,d_{\star\star}}\\
	0_{d_{\star\star}, d}&0_{d_{\star\star}, d}&0_{d_{\star\star},d_\star}&0_{d_{\star\star},d_\star}&0_{d_{\star\star},d^2}&0_{d_{\star\star}}
	\end{pmatrix}\,,\]
	with eigenvalues $0$, $ε>0$ and $γ>0$ and therefore fulfils the conditions for a Fuchsian system for all allowed decay coefficients on the right-hand side.
	
	To show the system (\ref{my_diff_evo_eqs}) is Fuchsian, each term on the right-hand side has to be shown to be exponentially decreasing.
	
	To simplify the argument the decay rates as $τ\to \infty$ of the coefficients $c_A$ and their derivatives, with the ansatz (\ref{myansatz}) for $\N{^a_i}$, are now listed:
	\begin{equation}\label{ca_decay}\begin{aligned}
	&c_{sym}=(P\indices{^j_a}\N{^b_j})^2/2=O(1)\,,
	&&\frac{∂c_{sym}}{∂P}=O(1)\,,
	&&\frac{∂c_{sym}}{∂\N{^a_i}}=O(1)\,,\\
	&\frac{∂c_{sym}}{∂(∂_j\N{^a_i})}=0\,,
	&&\frac{∂c_{sym}}{∂(∂_j∂_k\N{^a_i})}=0\,,\\
	&c_{d.g.}=(C\indices{^a_b_c})^2\propto(\N{_{,i}})^2=O(e^{-2γτ})\,,
	&&\frac{∂c_{d.g.}}{∂\N{^a_i}}=O(e^{-2γτ})\,,
	&&\frac{∂c_{d.g.}}{∂P\indices{^i_a}}=0\,,\\
	&\frac{∂c_{d.g.}}{∂(∂_j\N{^a_i})}\propto \N{_{,j}}=O(e^{-γτ})\,,
	&&\frac{∂c_{d.g.}}{∂(∂_j∂_k\N{^a_i})}=0\,,\\
	&c_{s.d.g.}=O(τ^2)\,,
	&&\frac{∂c_{s.d.g.}}{∂\N{^a_i}}\propto β^a_{,i}\N{_{,j}}=O(τe^{-γτ})\,,
	&&\frac{∂c_{s.d.g.}}{∂P\indices{^i_a}}=0\,,\\
	&\frac{∂c_{s.d.g}}{∂(∂_j\N{^a_i})}=O(τ)\,,
	&&\frac{∂c_{s.d.g}}{∂(∂_j∂_k\N{^a_i})}=O(1)\,.
	\end{aligned}\end{equation}
	
	Here $c_{sym}$ stands for the symmetry wall coefficients $(P\indices{^j_a}\N{^b_j})^2/2$, $c_{d.g.}$ for the dominant gravitational ones, $(C\indices{^a_b_c})^2/4$, and $c_{s.d.g.}$ for the subdominant gravitational ones defined in \eqref{subdom_grav_coeff}.
	
	As the $\bar{β}^a$ equation (\ref{my_beta_eq}) has no terms on the right-hand side we consider first the $\bar{π}_a$ equation (\ref{my_pi_eq}). The key terms are the exponentials $\exp(-2ω_A(β))$ and the $c_A$ and their derivatives. The overall space derivatives in the second and third part are innocuous as they can only bring down polynomial expressions in $τ$ from the exponentials.
	
	Beginning with the (relevant terms of the) first part, $\sum_A c_A e^{-2w_A(β_{[0]})}$, we look at the three kinds of walls (symmetry, dominant gravitational and subdominant gravitational) separately:
	\begin{description}
		\item[symmetry] The coefficients do not decay (see (\ref{ca_decay})). The whole term decays exponentially only if
		\begin{equation}\label{my_sym_cond}
		ω_{\text{sym }ab}(p_\circ)>ε \quad∀ a>b\,,
		\end{equation}
		\ie, as $ε$ is arbitrarily small, if the symmetry wall conditions are fulfilled (this means that $p_\circ^1<p_\circ^2<\dots<p_\circ^d$).
		\item[dom. grav.] The coefficients decay as $e^{-2γτ}$. The whole term shows exponential decay if
		\begin{equation}\label{my_dom_cond_1}
		-2γ-2α_{abc}(p_\circ)+ε<0 \Leftrightarrow γ>-α_{abc}(p_\circ)+ε\quad ∀ a,b,c\,,
		\end{equation}
		where the $α_{abc}$ are the linear forms of the dominant gravitational wall, defined in \eqref{dom_grav_walls}.
		\item[subdom. grav.] The coefficients do not decay, the whole term only decays if
		\begin{equation}\label{my_subdom_cond}
		μ_a(p_\circ)>ε\quad ∀ a\,,
		\end{equation}
		with $μ_a$ the linear forms of the subdominant gravitational walls, defined in \eqref{subdom_grav_walls}.
	\end{description}
	
	The second and third parts of the $\bar{π}_a$ equation \eqref{my_pi_eq} include only the subdominant gravitational walls, as these are the only ones containing derivatives of $β$. To guarantee exponential decay in these terms, the subdominant wall conditions $μ_a(p_\circ)>ε$ have to be fulfilled, as in (\ref{my_subdom_cond}).
	
	The $\N{_s^a_i}$ equation, \eqref{my_n_eq}, contains only the symmetry wall term, as the $P\indices{^i_a}$ derivative of the other coefficients vanishes. Because of the exponentially increasing term $\exp(γτ)$, decay requires that
	\begin{equation}\label{my_sym_cond_gamma}
	2ω_{\text{sym }ab}(p_\circ)>γ\quad ∀ a>b\,.
	\end{equation}
	
	The first part of the $\bar{P}\indices{^i_a}$ equation \eqref{my_p_eq} (containing $\N{^a_i}$ derivatives of the coefficients) decays exponentially if the following conditions are satisfied for the different wall types:
	\begin{description}
		\item[symmetry] The symmetry wall conditions  $ω_{\text{sym }ab}(p_\circ)>0$ have to be fulfilled (as in (\ref{my_sym_cond})).
		\item[dom. grav.] $γ>-α_{abc}(p_\circ)\quad ∀ a,b,c$ (as in (\ref{my_dom_cond_1})).
		\item[subdom. grav.] $μ_a(p_\circ)>0\quad ∀ a$ (as in (\ref{my_subdom_cond})).
	\end{description}
	The second and third part of the $\bar{P}\indices{^i_a}$ equation involve only the gravitational walls, as the symmetry wall coefficients don't contain any spatial derivatives of $\N{^a_i}$. Both the dominant and subdominant gravitational wall coefficients include first order (spatial) derivatives of $\N{^a_i}$, but only the subdominant walls contain second order derivatives. The dominant gravitational wall term gives the condition
	\begin{equation}\label{my_dom_cond_2}
	γ>-2α_{abc}(p_\circ)\quad ∀ a,b,c
	\end{equation}
	($2$ because the decay of the derivative of the coefficient is only of order $O(\exp(-γτ))$ instead of $O(\exp(-2γτ))$). The subdominant wall terms require $μ_a(p_\circ)>0$, as before.
	
	The $N\indices{_s^a_i_j}$ equation \eqref{my_N_eq} requires the same conditions as the $\N{_s^a_i}$ equation \eqref{my_n_eq}, as the overall space derivative only adds polynomial terms.
	
	Decay of the right-hand side of the $B^a_j$ equation \eqref{my_Bt_eq} requires only $ε>0$.
	
	Summarising, the conditions are
	\begin{equation}\label{my_conds}
	\begin{alignedat}{7}
	ω_{\text{sym }ab}(p_\circ)&>ε\,, &\qquad&γ+α_{abc}(p_\circ)&>ε\,, &\qquad& μ_a(p_\circ)&>ε\,,\\
	2ω_{\text{sym }ab}(p_\circ)&>γ\,, &\qquad& γ+2α_{abc}(p_\circ)&>0\,,&\qquad & ε&>0\,,
	\end{alignedat}
	\end{equation}
	for all indices $a,b,c$ and the asymptotic Hamiltonian constraint
	\begin{equation}\label{my_ham_const}
	\sum_{a\neq b}p_\circ^a p_\circ^b=0\,.
	\end{equation}
	Additionally, the asymptotic momentum constraint equation has to be fulfilled. This will be discussed in detail later.
	
	\subsection{$d=3$ case}
	In $3+1$ dimensions the conditions (\ref{my_conds}) are explicitly (without removing redundant ones)
	\begin{subequations}\label{my_d3_conds}
		\begin{align}
		ω_{sym}(p_\circ)&>ε&\Rightarrow&&p_\circ^3-p_\circ^2&>ε\,,&p_\circ^3-p_\circ^1&>ε\,,&p_\circ^2-p_\circ^1&>ε\,,\label{symcond1}\\
		γ+α_{abc}(p_\circ)&>ε&\Rightarrow&&γ+2p_\circ^1&>ε\,,&γ+2p_\circ^2&>ε\,,&γ+2p_\circ^3&>ε\,,\label{dgcond1}\\
		μ_a(p_\circ)&>ε&\Rightarrow&&p_\circ^1+p_\circ^2&>ε\,,&p_\circ^1+p_\circ^3&>ε\,,&p_\circ^2+p_\circ^3&>ε\,,\label{sdgcond}\\
		2ω_{sym}(p_\circ)&>γ&\Rightarrow&&2(p_\circ^3-p_\circ^2)&>γ\,,&2(p_\circ^3-p_\circ^1)&>γ\,,&2(p_\circ^2-p_\circ^1)&>γ\,,\label{symcond2}\\
		γ+2α_{abc}(p_\circ)&>0&\Rightarrow&&γ+4p_\circ^1&>0\,,&γ+4p_\circ^2&>0\,,&γ+4p_\circ^3&>0\,,\label{dgcond2}\\
		&&&& ε&>0\,,
		\end{align}
	\end{subequations}
	and the asymptotic Hamiltonian constraint is
	\begin{equation}\label{my_d3_ham_const}
	p_\circ^1p_\circ^2+p_\circ^1p_\circ^3+p_\circ^3p_\circ^2=0\,.
	\end{equation}
	Let us show that they can be fulfilled simultaneously:
	
	The Hamiltonian constraint (\ref{my_d3_ham_const}) gives
	\begin{equation}\label{p1exp}
	p_\circ^1=-\frac{p_\circ^2p_\circ^3}{p_\circ^2+p_\circ^3}\,.
	\end{equation}
	The conditions (\ref{symcond1}) follow from (\ref{symcond2}) by choosing $ε<γ/2$. Likewise, (\ref{dgcond1}) follows from (\ref{dgcond2}) by choosing $ε<-2p_\circ^1$ (as $γ>0$ and $p_\circ^1<0$). The second and third condition in (\ref{sdgcond}) follow from the first and from $p_\circ^3>p_\circ^2>0>p_\circ^1$ (from (\ref{symcond1})), as does the first when inserting (\ref{p1exp}) and choosing $ε$ sufficiently small. The second condition in (\ref{symcond2}) follows from the first. The remaining ones are now
	\[
	\min\left\{2(p_\circ^3-p_\circ^2),2(p_\circ^2-p_\circ^1)\right\}>γ>-4p_\circ^1\,,
	\]
	(the last two conditions in (\ref{dgcond2}) follow from $p_\circ^3>p_\circ^2>0$ and $γ>0$). $2(p_\circ^2-p_\circ^1)>-4p_\circ^1$ follows from $p_\circ^2>0$ after inserting (\ref{p1exp}). The last remaining condition, $2(p_\circ^3-p_\circ^2)>-4p_\circ^1$, gives $(\sqrt{2}-1)p_\circ^3>p_\circ^2$.
	
	Summarizing, the conditions on $p_\circ^a$ (at each spatial point) are
	\begin{equation}\label{final_my_p_cond}
	p_\circ^3>0\,,\qquad 0<p_\circ^2<(\sqrt{2}-1)p_\circ^3\quad\text{and}\quad p_\circ^1=-\frac{p_\circ^2 p_\circ^3}{p_\circ^2+p_\circ^3}\,.
	\end{equation}
	The remaining free parameters are the six functions of the space coordinates $P\indices{_\circ^i_a}$ and $β_\circ^a$.
	
	\subsection{Constraints}
	
	In addition to the evolution equations, the constraints \eqref{full_ham_const} and \eqref{full_mom_const} also have to be fulfilled. The asymptotic Hamiltonian constraint was already included in the conditions discussed in the last section. We start by considering the asymptotic momentum constraints and then show that the full constraints are satisfied if the asymptotic ones are.
	
	Inserting the ansatz \eqref{myansatz} (with $\N{_\circ}=\1$) into the asymptotic momentum constraint \eqref{asym_mom_const} gives
	\begin{equation}\label{my_asym_mom_const}
	\frac{1}{2}\sum_{\substack{b \\ b>a}}P\indices{_\circ^b_a_{,b}}-\left(G_{ac}p_{\circ ,a}^c +β^d_{\circ ,a} p_\circ^f G_{df}\right)=0\qquad∀a\,,
	\end{equation}
	(all terms containing $C\indices{_{[0]}^a_b_c}$ vanish as the asymptotic limit of $\N{}$ is constant in space).
	
	This is equivalent to
	\begin{align}
	\nonumber
	β^1_{\circ,3}
	&=-(p_\circ^2+p_\circ^3)^{-1}(p_{\circ,3}^2+p_{\circ,3}^1+β_{\circ,3}^2(p_\circ^1
	+p_\circ^3)+β^3_{\circ,3}(p_\circ^1+p_\circ^2))\,,
	\\
	\nonumber
	\p{_\circ^3_2_{,3}}&=2\left(G_{2c}p_{\circ,2}^c+β_{\circ,2}^d p_\circ^f G_{df}\right)\,,
	\\
	\label{28X14.1}
	\p{_\circ^3_1_{,3}}&=-\p{_\circ^2_1_{,2}}+2\left(G_{1c}p_{\circ,1}^c+β^d_{\circ,1}p_\circ^f G_{df}\right)
	\,.
	\end{align}
	Given any functions $β^2_\circ$, $β^3_\circ$,  $\p{_\circ^2_1}$ and $p_\circ^1$, $p_\circ^2$, $p_\circ^3$ fulfilling \eqref{final_my_p_cond} one can determine $ β^1_{\circ} $, $\p{_\circ^3_2}$ and $\p{_\circ^3_1}$ from \eqref{28X14.1}, obtaining thus a solution of the asymptotic constraint equations.
	
	The full constraints are fulfilled if the asympotic ones are, following the arguments of section \ref{sec:asym_full_const}. The condition that the (modified) evolution equations for the constraints \eqref{asym_full_fuchs} are of Fuchsian form requires that the subdominant gravitational wall conditions are satisfied, which is the case here (see \eqref{my_conds}). To ensure the full momentum constraints converge to the asymptotic ones the symmetry wall conditions have to be fulfilled, which is also the case. Finally, the Hamiltonian converges to the asymptotic Hamiltonian if all terms after the first one in \eqref{ham_iwa_full} vanish asymptotically. For the terms coming from the symmetry and subdominant gravitational walls this is ensured by the decay of the exponential terms, as for the general ansatz. For the dominant gravitational wall terms it follows from the decay of the coefficients, which contain spatial derivatives of $\N{^a_i}$, and the resulting inequalities \eqref{dgcond2}.
	
	\section{Analysis of the constructed solutions}\label{sec:analysis}
	In the following we will need the asymptotic behaviour of the metric components of the constructed solutions. This is derived in \ref{sec:asym_beh}.
	
	From the results above
	there exist numbers $\gamma>0$ and $\nu>0$ so that
	\begin{equation}
	\beta^a  = p^a_{\circ} \tau +  \beta^a_{\circ} + O(e^{-\nu \tau})
	\;,
	\quad
	\N{^a_i}= \delta^a_i+ O(e^{-(\gamma+\nu) \tau})
	\;.
	\label{9XII14.5}
	\end{equation}
	Setting
	\newcommand{\sigmap}{\sigma_{p_\circ} }
	\newcommand{\sigmab}{\sigma_{\beta_\circ} }
	\begin{equation}
	\sigmap = p^1_{\circ} + p^2_{\circ}  + p^3_{\circ}  >0
	\;,
	\quad
	\sigmab = \beta^1_{\circ} + \beta^2_{\circ}  + \beta^3_{\circ}
	\;,
	\label{9XII14.6}
	\end{equation}
	the leading order behaviour of the metric components, and those of its inverse, is
	\begin{align*}
	\bar{g}_{00} & =  -e^{-2\sigmap τ - 2 \sigmab }(1 + O^ν)\to0
	\,,&
	\bar{g}^{00}&=-e^{2\sigmap τ + 2 \sigmab }(1 + O^ν)
	\,,
	\\
	\bar{g}_{0 i} & \equiv  0
	\,,&
	\bar{g}^{0 i} & \equiv 0
	\,,
	\\
	\bar{g}_{ii} & = e^{-2p^i_{\circ} \tau -2  \beta^i_{\circ}  }(1 + O^ν)
	\,,&
	\bar{g}^{ii} & = e^{2p^i_\circ τ + 2 β^i_\circ}(1 + O^ν)
	\,,
	\\
	\bar{g}_{12} & = e^{-2p_\circ^2τ}e^{-2β_\circ^1}(K\indices{^1_2}+O^ν)\to0
	\,,&
	\bar{g}^{12} & = -e^{2p_\circ^1τ}e^{2β_\circ^2}(K\indices{^1_2}+O^ν)\to 0
	\,,
	\\
	\bar{g}_{13} & = e^{-2p_\circ^3τ}e^{-2β_\circ^1}(K\indices{^1_3}+O^ν)\to 0
	\,,&
	\bar{g}^{13} & = -e^{2p_\circ^1τ}e^{2β_\circ^3}(K\indices{^1_3}+O^ν)\to 0
	\,,
	\\
	\bar{g}_{23} & = e^{-2p_\circ^3τ}e^{-2β_\circ^2}(K\indices{^2_3}+O^ν)\to 0
	\,,&
	\bar{g}^{23} & = -e^{2p_\circ^2τ}e^{2β_\circ^3}(K\indices{^2_3}+O^ν)
	\,,
	\end{align*}
	where $O^ν:=O(e^{-ντ})$, the $K^a{}_i$ are functions depending only on the spatial coordinates and with the behaviour above preserved under differentiation in the obvious way. The precise form of the $K^a{}_i$ is given in \ref{sec:asym_beh}.
	
	\subsection{Remaining coordinate freedom}\label{sec:coord_free}
	
	We wish, now, to analyse how the ansatz \eqref{myansatz}, with the choice $\N{_\circ}=\1$, constrains the remaining coordinate freedom.
	
	In \cite{Damour2003} it is asserted that transformations mixing time and space coordinates are prohibited by the choice of lapse and shift and the assumption that the singularity is approached as $τ\to\infty$.
	Presumably this should follow from the resulting equations
	\begin{align}
	g_{ij}\frac{∂x^i}{∂\tilde{τ}}\frac{∂x^j}{∂y^k}&=\det g \frac{∂τ}{∂\tilde{τ}}\frac{∂τ}{∂y^k}
	\;,
	\label{coord_shift}
	\\
	-\det g \left(\frac{∂τ}{∂\tilde{τ}}\right)^2+g_{ij}\frac{∂x^i}{∂\tilde{τ}}\frac{∂x^j}{∂\tilde{τ}}&=-\det\left(-\det g\left(\frac{∂τ}{∂y^k}\right)^2δ_{kl}+g_{ij}\frac{∂x^i}{∂y^k}\frac{∂x^j}{∂y^l}\right)
	\,,
	\label{coord_lapse}
	\end{align}
	(assuming a transformation $τ,x^i\to \tilde{τ}(τ,x^j),y^i(τ, x^j)$).
	However, the assertion is not clear.
	%\ptcr{I understand that you have tried to derive a Fuchsian system of equations from the above, without success, should be said. Actuallly it would really be useful if you shortly describe  what you did and why this did not work}
	%\pk{Added}
	
	%\ptcr{check if the system below can be written in Fuchsian form for the variables $(\tilde x^s, A^i{}_j$ and show uniqueness in the relevant class of coordinate transformations if true}
	%
	An attempt was made to construct a Fuchsian system by starting from the transformation law of the Christoffels. Defining
	\[
	A^α_β:=\frac{∂ y^α}{∂x^β}\,,
	\]
	and writing the transformation in terms of it gives
	\begin{equation}\label{16III15.2}
	\frac{\partial A ^α{}_β}{ \partial{x^γ}}
	=   A^δ{}_γ A^ε{}_β {\tilde
		\Gamma}^α{}_{δ ε} -A^α{}_δ \Gamma^δ{}_{γβ}
	\,.
	\end{equation}
	%
	%with the asymptotic condition that $\tilde x^0\to_{\tau\to \infty}\infty$.
	We split $A^α{}_β$ into $A^α{}_β=A_\circ{}^α{}_β+δA^α{}_β$ with the assumptions $A_\circ{}^0{}_β=A_\circ{}^α{}_0=0$, $A_\circ{}^a{}_b=A_\circ{}^a{}_b(x^i)$ and $A_\circ{}^0{}_0=A_\circ{}^0{}_0(τ)$. Inserting this into the $γ=0$ equation of \eqref{16III15.2} gives
	\begin{equation}
	∂_τδA^α{}_β=-δ^α_τδ^τ_β∂_τA_\circ{}^τ{}_τ+(δA^δ{}_τ\tilde{Γ}^α{}_{δε}+A_\circ{}^τ{}_τ \tilde{Γ}^i{}_{τε})(A_\circ{}^ε{}_β+δA^ε{}_β)-(A_\circ{}^α{}_δ+δA^α{}_δ)Γ^δ{}_{τβ}\,.
	\end{equation}
	This system is unfortunately not of Fuchsian form, as many of the $Γ^α{}_{βγ}$ diverge as $τ\to\infty$ (see \ref{sec:asym_beh}).
	
	Assuming nevertheless a transformation of the form
	\[
	(τ, x^i)\to (\tilde{τ}(τ, x^i), y^i(x^j))\,,
	\]
	and inserting into \eqref{coord_shift} gives $∂τ/∂y^k=0$ and therefore $\tilde{τ}=\tilde{τ}(τ)$. \eqref{coord_lapse} then leads to
	\[
	\left(\frac{∂τ}{∂\tilde{τ}}\right)^2=\det\left(\frac{∂x^i}{∂y^j}\right)^2\,,
	\]
	which implies that the Jacobi determinant of the spatial transformation is constant (in space and time) and that $\tilde{τ}$ is an affine
	function of $τ$.
	
	%\pk{Rewritten from here\dots}
	Starting with a general spatial coordinate transformation $x^i\to y^i(x^j)$ the metric components $\tilde{g}_{kl}$ in the new frame take the form
	\begin{equation}\label{24III15.0}
	\tilde{g}_{kl}=\sum_a e^{-2β^a}\N{^a_i}\N{^a_j}\frac{∂x^i}{∂y^k}\frac{∂x^j}{∂y^l}
	=\sum_ae^{-2\tilde{β}^a}\tilde{\mathcal{N}}\indices{^a_k}\tilde{\mathcal{N}}\indices{^a_l}\,.
	\end{equation}
	For $\tilde{g}_{11}$ this gives
	\begin{equation}\label{24III15.1}
	\tilde{g}_{11}=e^{-2β^1}\left(\frac{∂x^1}{∂y^1}\right)^2+O(X)=e^{-2\tilde{β}^1}\,,
	\end{equation}
	with $O(X)=O(\exp(-2p_\circ^2τ))$ denoting higher order terms. Assuming $∂x^i/∂y^i\neq0$, which is necessary to preserve the conditions $p_\circ^3>p_\circ^2>p_\circ^1$ \eqref{symcond1}, this implies $\tilde{β}^1=β^1-\log(∂x^1/\,∂y^1)$.
	
	The equation for $\tilde{g}_{12}$ is
	\begin{equation}\label{24III15.2}
	\tilde{g}_{12}=e^{-2β^1}\frac{∂x^1}{∂y^1}\frac{∂x^1}{∂y^2}+O(X)=e^{-2\tilde{β}^1}\tilde{\mathcal{N}}\indices{^1_2}\,.
	\end{equation}
	Inserting $\exp(-2\tilde{β}^1)$ from \eqref{24III15.1} yields
	\begin{equation}\label{24III15.3}
	\tilde{\mathcal N}\indices{^1_2}=\frac{∂x^1/\,∂y^2}{∂x^1/\,∂y^1}+O(X)\,.
	\end{equation}
	Requiring $\tilde{\mathcal N}\indices{^a_i}=δ^a_i+O(\exp(-\tilde{γ}τ))$ we obtain the condition $∂x^1/∂y^2=0$.
	Similarly, from the $\tilde{g}_{13}$ equation of \eqref{24III15.0}, we get
	\begin{equation}\label{24III15.4}
	\tilde{\mathcal N}\indices{^1_3}=\frac{∂x^1/\,∂y^3}{∂x^1/\,∂y^1}+O(X)\,,
	\end{equation}
	and therefore $∂x^1/\,∂y^3=0$.
	
	Using these conditions on $x^1$, the $\tilde{g}_{22}$ equation becomes
	\begin{equation}\label{24III15.5}
	\tilde{g}_{22}=e^{-2β^2}\left(\frac{∂x^2}{∂y^2}\right)^2+O(\tilde{X})=e^{-2\tilde{β}^2}+(\tilde{\mathcal N}\indices{^1_2})^2e^{-2\tilde{β}^1}\,,
	\end{equation}
	where $O(\tilde{X})=O(\exp(-2p_\circ^3τ))$ denotes higher order terms and therefore $\tilde{β}^2=β^2-\log(∂x^2/\,∂y^2)$.
	
	The $\tilde{g}_{23}$ equation is
	\begin{equation}\label{24III15.6}
	\tilde{g}_{23}=e^{-2β^2}\frac{∂x^2}{∂y^2}\frac{∂x^2}{∂y^3}+O(\tilde{X})=\underbrace{\tilde{\mathcal N}\indices{^1_2}\tilde{\mathcal N}\indices{^1_3}e^{-2\tilde{β}^1}}_{=O(\tilde{X})}+\tilde{\mathcal N}\indices{^2_3}e^{-2\tilde{β}^2}
	\end{equation}
	and yields
	\begin{equation}
	\tilde{\mathcal N}\indices{^2_3}=\frac{∂x^2/\,∂y^3}{∂x^2/\,∂y^2}+O(X)
	\end{equation}
	which implies $∂x^2/\,∂y^3=0$.
	
	The conditions on the coordinate transformation are now
	\begin{equation}\label{19III15.1}\begin{aligned}
	x^1&=x^1(y^1)\,,\\
	x^2&=x^2(y^1,y^2)\,,\\
	x^3&=x^3(y^1,y^2,y^3)\,.
	\end{aligned}\end{equation}
	Under such a coordinate change the asymptotic functions $\N{_\circ^a_i}$, $p_\circ^a$ and $β_\circ^a$ transform as
	\begin{equation}\label{24III15.7}
	\tilde{\mathcal N}\indices{_\circ^a_i}=\N{_\circ^a_i}=δ^a_i\,,\qquad
	\tilde{p}_\circ^a = p_\circ^a\,,\qquad
	\tilde{β}_\circ^a = β_\circ^a-\log\left(\frac{∂x^a}{∂y^a}\right)\,.
	\end{equation}
	As the $p_\circ^a$ remain unchanged, the conditions \eqref{final_my_p_cond} are unaffected. Therefore $γ$ can be chosen to have the same value in the new coordinates, \ie $\tilde{γ}=γ$.
	
	The transformations \eqref{24III15.7} reduce the possible isometries of the constructed solutions: Spatial isometries would have to be of the form \eqref{19III15.1} as otherwise the asymptotic evolutions would not match (a $\tilde{\mathcal N}_\circ\neq \1$ would not give an asymptotically diagonal metric and transformations which exchange the order of the $β^a$ would change the asymptotics of the diagonal terms).
	As $p_\circ^2$ and $p_\circ^3$ are only constrained by the inequalities $0<p_\circ^2<(\sqrt{2}-1)p_\circ^3$ but otherwise free functions of all space coordinates, which are not influenced by coordinate transformations of this form, we have the following proposition:
	
	\begin{prop}
		For a generic choice of the asymptotic functions $p_\circ^2$ and $p_\circ^3$, the corresponding solutions have no (continuous or discrete) isometries $ϕ:M\to M$ of the form $τ(ϕ(q))=τ(q),\; ∀ q\in M$, \ie involving only the spatial coordinates.
	\end{prop}
	
	% \ptcr{there should be a conclusion in the form of a Proposition: if ... then there are no isometries of the solution satisfying ...; a distinction between discrete and continuous isometries should be made clear, which justifies the calculations in the next section}
	%\ptcr{it would be good to write here the transformation laws of the different important functions under the above coordinate changes, and draw a conclusion about isometries}\pk{\dots to here}
	
	\subsection{Killing vectors}\label{sec:killing}
	
	In this section we wish to investigate continuous symmetries of the constructed solutions. This requires an analysis of the Killing equations:
	\begin{equation}
	\label{9XII14.1}
	\mcL_Xg_{\mu\nu} = X^\sigma \partial_\sigma g_{\mu\nu} + \partial_\mu X^\sigma g_{\sigma\nu}
	+ \partial_\nu X^\sigma g_{\sigma\mu}
	\;.
	\end{equation}

	We will seek Killing vectors of the form $X=X^τ∂_τ+X^i∂_i$, with
	\begin{equation}\label{9XII14.11}\begin{split}
	X^τ &= O(e^{-\nu\tau})
	\,,
	\end{split}\end{equation}
	and we will assume that the behaviour above is also preserved under differentiation. This ansatz is more general than the assumption of purely spatial isometries in the previous section, which would imply $X^τ=0$. It does, however, only include isometries which can be described by Killing vectors, \ie continuous but not discrete ones.
	% \ptcr{mention that this is actually an ansatz more general than that of the previous section, which would require $X^\tau =0$, within the setting of continuous isometries}\pk{added}
	
	The $τi$ killing equation states
	\[
	\mcL_X g_{\tau i}  =   \partial_\tau X^j g_{j  i}
	+ \partial_i  X^\tau g_{\tau\tau}=0.
	\]
	Contracting with $g^{ik}$ gives\vspace{-1.5em}
	\[
	∂_τX^k=-g^{ik}g_{ττ}∂_iX^τ=O(e^{-\big(\overbrace{2p_\circ^2-2p_\circ^1}^{>0}+ν\big)τ})=O(e^{-ντ})\,,
	\]
	\ie the $τ$ derivative of the spatial components of the Killing field decay exponentially.
	% $ and $\partial_\sigma X^ i= O(1)$.
	% \ptcr{one should be able to prove this? probably follows from the previous section either as a hypothesis or a consequence?}
	
	Considering the $ττ$ component of the Killing equation gives
	\begin{equation}\begin{split}
	\mcL_X g_{\tau\tau} & = X^i ∂_i g_{ττ} +X^τ∂_τ g_{ττ}+ 2g_{ττ} ∂_τX^τ\\
	& =  2 e^{-2\sigmap \tau - 2 \sigmab }\bigg[   X^i \partial_i  \left(\sigmap \tau + \sigmab\right)+\underbrace{σ_{p_\circ}X^τ-∂_τX^τ}_{=O(e^{-ντ})}\bigg]\left(1 + O^ν\right)
	\;.
	\end{split}\end{equation}
	The term of order one inside the bracket is non-zero for large times unless
	\begin{equation}\label{5I15.1}
	X^i_{\circ} \partial_i  \sigmap =0
	\ \mbox{  and }
	\ X^i _{\circ}\partial_i  \sigmab =0
	\;.
	\end{equation}
	The $ii$ component of the Killing equation gives
	\begin{equation}\begin{split}
	\mcL_X g_{ii}=&X^k ∂_kg_{ii}+X^τ∂_τg_{ii}+2∂_iX^k g_{ki}=\\
	&e^{-2(p_\circ^iτ+β_\circ^i)}(1+O^ν)\bigg[-2X_\circ^k ∂_k(p_\circ^i τ+β_\circ^i)-\smash{\underbrace{2X^τp_\circ^i}_{=O(e^{-ντ})}}\bigg]\\
	&+2∂_iX_\circ^iO(e^{-2(p_\circ^iτ+β_\circ^i)}) \,,
	\end{split}
	\end{equation}
	which is certainly non-zero unless the highest order term, containing $p_\circ^i τ$, vanishes. This requires
	\begin{equation}\label{5I15.2}
	X_\circ^k ∂_k p_\circ^i=0\,.
	\end{equation}
	There are no solutions $X^μ$ fulfilling these conditions in general: The second equation of \eqref{5I15.1} and \eqref{5I15.2} can be combined into $A\indices{^i_k} X^k=0$, with $A$ containing the derivatives of $p_\circ^2$, $p_\circ^3$ and $σ_{β_\circ}$. If the determinant of $A$ is non-zero the only solution is $X_\circ^k= 0$. If this holds in the neighbourhood of a point, the Killing vector vanishes everywhere. We conclude that our solutions will not have any Killing vectors of the form \eqref{9XII14.11} in general.
	\begin{prop}
		For a generic choice of the asymptotic functions $p_\circ^2$, $p_\circ^3$ and $β_\circ^a$ the corresponding solutions contain no Killing vectors $X=X^τ∂_τ+X^i∂_i$ satisfying $X^τ=O(e^{-ετ})$ for any $ε>0$.
	\end{prop}

	Considering now a general Killing vector field $X^μ$, satisfying
	\begin{equation}\label{16III15.3}
	\nabla_μ X_ν+\nabla_ν X_μ=0\,,
	\end{equation}
	we start with the expression $\nabla_τ\nabla_αX_β$. Using the definition of the Riemann tensor we obtain
	\[
	\nabla_τ\nabla_α X_β=g_{βγ}\nabla_τ\nabla_α X^γ=\nabla_α\nabla_τ X_γ + R_{βγτα}X^α\,.
	\]
	Using the antisymmetry of the Riemann tensor and the first Bianchi identity gives
	\[
	\nabla_τ\nabla_α X_β=\nabla_α\nabla_τ X_γ-R_{γβτα}X^γ=\nabla_α\nabla_τ X_γ+(R_{γταβ}+R_{γαβτ})X^γ\,.
	\]
	Applying \eqref{16III15.3} and using the definition of the Riemann tensor again leads to
	\begin{align*}
	\nabla_τ\nabla_α X_β&=R_{γταβ}X^γ-\nabla_α\nabla_βX_τ\overbrace{-\nabla_β\nabla_τX_α+\nabla_τ\nabla_βX_α}^{R_{γαβτ}X^γ}\\
	&=R_{γταβ}X^γ-\nabla_α\nabla_βX_τ+\nabla_β\nabla_αX_τ-\nabla_τ\nabla_αX_β\\
	&=2R_{γταβ}X^γ-\nabla_τ\nabla_αX_β\,,
	\end{align*}
	and therefore
	\[
	\nabla_τ\nabla_α X_β=R_{γταβ}X^γ\,.
	\]
	Introducing
	\[
	F_{αβ} = \frac{1}{2} (\nabla_α X_β - \nabla_β X_α)=\nabla_α X_β
	\,,
	\]
	we thus have the following system of equations for the pair $(X,F)$ if $X$ is a Killing vector:
	\begin{equation}\label{13I15.1}
	\left\{
	\begin{aligned}
	\nabla_τ X^λ &= g^{λi}F_{τ i} \,, \\
	\nabla_τ F_{αβ} &= R_{σταβ} X^σ\,.
	\end{aligned}
	\right.
	\end{equation}
	This has the general structure of a Fuchsian system, but does not fulfil the conditions given in section \ref{sec:fuchsthm}. By redefining some of the $X^α$ and $F_{αβ}$ the equations can be brought to a form where all terms on the right-hand side decay exponentially, but the equation for $X^0$ contains the term $Γ^0_{00}X^0$ on the left-hand side, which gives a negative eigenvalue $-(p_\circ^1+p_\circ^2+p_\circ^3)$ in the matrix $A$. To still satisfy the conditions of the Fuchs theorem all terms would have to decay faster than $\exp(-τ(p_\circ^1+p_\circ^2+p_\circ^3))$, which is not possible.
	% The equation for $X^2$ is
	%\begin{equation}\label{16III15.4}
	%∂_τX^2=g^{2λ}F_{0λ}-Γ^2{}_{0λ}X^λ\,.
	%\end{equation}
	%The first term on the right-hand side contains $g^{22}F_{02}=O(\exp(2p_\circ^2τ))F_{02}$ which diverges as $τ\to\infty$. This can be compensated by defining a new variable $\bar{F}_{02}=\exp(2p_\circ^2)F_{02}$ which turns it into a term of order $1$. However the new equation for $\bar{F}_{02}$ is then given by
	%\begin{equation}\label{16III15.5}
	%∂_τ\bar{F}_{02}=2p_\circ^2\bar{F}_{02}+e^{2p_\circ^2τ}\left(R_{λ002}X^λ+Γ^λ{}_{00}F_{λ2}+Γ^λ{}_{02}F_{0λ}\right)\,,
	%\end{equation}
	%with the curvature term containing $\exp(2p_\circ^2τ)R_{2002}X^2$. As the $R_{2002}$ component of the Riemann tensor is of order $O(τ\exp(-2p^2τ))$, this expression is $O(τ)$, violating the conditions of the Fuchs theorem. Redefining $X^2$ cannot change this, as any change would also affect the $X^2$ equation \eqref{16III15.4} requiring a further redefinition of $\bar{F}_{02}$.
	%The equation for $F_{12}$ contains the term $R_{1012}X^1$ which behaves as $(C+O(e^{-ντ}))τe^{-2p_\circ^1τ}X^1$. The divergence can be compensated by replacing $X^1$ with $\bar{X}^1=X^1e^{-2p_\circ^1τ+ετ}$. However the new equation
	
	\subsection{Relationship with previously known solutions}\label{sec:relprev}
	The first class of vacuum spacetimes for which asymptotically simple behaviour was shown was the polarized Gowdy class \cite{Isenberg:1989gq,Chrusciel:1990wn}. This is a class of solutions containing two commuting spacelike Killing vector fields (the polarization condition) with constant $t$ hypersurfaces which are compact without boundary and orientable (the Gowdy condition). The topology of spacelike slices of these spacetimes is constrained to one of $T^3$, $S^2\times S^1$, $S^3$ or a Lens space $L(p,q)$. In the following only the $T^3$ case will be considered. The metric for this case is given by
	\begin{equation}\label{pol_gowdy}
	ds^2=e^{2a}(-\dt^2+\Dθ^2)+t(e^{W}\dx^2+e^{-W}\dy^2)\,,
	\end{equation}
	where $a$ and $W$ are functions of $t$ and $θ$ which are $2π$ periodic in $x$.
	
	Redefining the $t$ coordinate as $t=e^{-τ}$ transforms the metric to
	\[
	\ds^2=-e^{2(a-τ)}\D τ^2+e^{2a}\Dθ^2+e^{W-τ}\dx^2+e^{-W-τ}\dy^2\,,
	\]
	which is directly in the gauge used here: The shift vanishes and the lapse $e^{2(a-τ)}$ is equal to the determinant of the spatial part of the metric. Comparing with \eqref{10II15.1} shows
	\begin{align*}
	e^{-2β^1}&=e^{2a}\Rightarrow β^1=-a\,,\\
	e^{-2β^2}&=e^{W-τ}\Rightarrow β^2=\frac{τ-W}{2}\,,\\
	e^{-2β^3}&=e^{-W-τ}\Rightarrow β^3=\frac{τ+W}{2}\,,\\
	\N{^1_2}&=\N{^2_3}=\N{^1_3}=0\,.
	\end{align*}
	
	The results of Chru\'sciel, Isenberg and Moncrief show that solutions of this form are parametrised by two functions of the $θ$ coordinate, $π$ and $ω$, appearing in the asymptotic expansion of $a$ and $W$ \cite{Isenberg:1989gq}. The expansion is
	\begin{align*}
	W&=π(τ-τ_0)+ω+O(τe^{-2τ})\,,\\
	a&=(1-π^2)(τ-τ_0)/4+α+O(τe^{-2τ})\,,
	\end{align*}
	where $τ_0$ is a constant and $α$ a function of $θ$ which can be determined from $π$ and $ω$. This gives for the $p_\circ^a$
	\begin{equation}\label{16III15.1}
	p_\circ^1=\frac{1}{4}(π^2-1)\,,\quad p_\circ^2=\frac{1}{2}(1-π)\,,\quad p_\circ^3=\frac{1}{2}(1+π)\,.
	\end{equation}
	
	These satisfy the asymptotic Hamiltonian constraint \eqref{asym_ham_const} but not necessarily the inequalities \eqref{final_my_p_cond}. These would imply $\sqrt{2}-1<π<1$ but there are solutions of the form \eqref{pol_gowdy} for any function $π$. The assumption that the metric coefficients only depend on the $θ$ space coordinate and $\N{^a_i}\equiv δ^a_i$ causes some of the potential walls to vanish: The coefficients of the dominant gravitational walls are proportional to $\N{^a_i_{,j}}$ and therefore vanish identically. For $\N{^a_i}$ to be constant, $\p{^i_a}$ has to vanish, which causes the symmetry walls to vanish. The coefficients of the subdominant gravitational walls \eqref{subdom_grav_coeff} contain terms which are not proportional to $\N{^a_i_{,j}}$. These do, however, contain spatial derivatives of the $β$, most of which are zero here. As $β$ only depends on $t$ and $θ$, only one of the walls, with linear form $μ_1(β)=β^2+β^3$, remains. Therefore, assuming that the metric coefficients depend only upon $θ$ and that $\N{^a_i}\equiv\1$, the only conditions left in \eqref{final_my_p_cond} are
	\begin{equation}\label{16III15.6}
	p_\circ^1=-\frac{p_\circ^2p_\circ^3}{p_\circ^2+p_\circ^3}
	\quad\text{and}\quad
	p_\circ^2+p_\circ^3>0\,.
	\end{equation}
	These conditions are satisfied by \eqref{16III15.1} which implies $p_\circ^2+p_\circ^3=1$ and $p_\circ^1=-p_\circ^2 p_\circ^3$.
	
	(One should note that not every solution satisfying \eqref{16III15.6} is of polarized Gowdy type, as $p_\circ^2$ and $p_\circ^3$ can still be independently specified.)
	
	More general (non-Gowdy) $T^2$ symmetric spacetimes have also been shown to exhibit simple asymptotic behaviour. These take the general form
	\[
	\ds^2=e^{2(η-U)}(-α\dt^2+\dx^2)+e^{2U}(\dy+A\D z+(G_1+A G_2)\dx)^2+e^{-2U}t^2(\D z+G_2\dx)^2\,,
	\]
	with $η$, $U$, $α$, $A$, $G_1$ and $G_2$ depending only on $t$ and $x$ \cite{Ames2013}. To obtain simple behaviour either polarization, corresponding to $A=\text{const}$, or half-polarization, corresponding to a restriction on the asymptotic behaviour of $A$, has to be assumed. In both cases the resulting spacetimes are not contained in the class constructed here. The functions $G_1$ and $G_2$ tend to constant (in $t$) functions of $x$, but they appear in the $\N{^a_i}$ in the Iwasawa decomposition. This conflicts with the assumption $\mathcal{N}\to\1$ (or $\to\text{const}$) made in constructing the new class. In this sense the new class is therefore more restricted than the polarized and half-polarized $T^2$ classes. However it includes free functions depending on all space coordinates, not just one.
	
	The Killing vectors of the Gowdy and general $T^2$-symmetric spacetimes are of the form considered in section \ref{sec:killing}, as they do not include derivatives with respect to $t$. These are therefore in general not present in the class of solutions constructed here.

	\section{Conclusion}\label{sec:conclusion}
	
	We have constructed a new class of four-dimensional (analytic) solutions to the vacuum Einstein equations which show asymptotically simple behaviour near a spacelike singularity, approached as $τ\to\infty$. The metric takes the form
	\begin{equation}\label{10II15.1}
	\ds^2=-e^{-2\sum_a β^a}\D τ^2+\sum_a e^{-2β^a}\N{^a_i}\N{^a_j}\dx^i\dx^j\,,
	\end{equation}
	with $β^a$ and $\N{^a_i}$ depending on all coordinates $τ$, $x^i$ and behaving asymptotically as
	\begin{equation}\label{10II15.2}
	β^a=β_\circ^a+τp_\circ^a+O(e^{-ντ})\quad\text{and}\quad
	\N{^a_i}=δ^a_i+O(e^{-(γ+ν)τ})\,,
	\end{equation}
	where $γ$ and $ν$ are positive constants.
	
	The class of solutions includes three completely free functions of all space coordinates, $β_\circ^2$, $β_\circ^3$, $P\indices{_\circ^2_1}$ ($P\indices{_\circ^2_1}$ does not appear in \eqref{10II15.1} and \eqref{10II15.2} but influences the exponentially decaying terms, as detailed in \ref{sec:asym_beh}) and two functions, $p_\circ^2$ and $p_\circ^3$, also depending on all space coordinates, which are constrained by the inequalities
	\[
	p_\circ^3>0\,,\qquad 0<p_\circ^2<(\sqrt{2}-1)p_\circ^3\,.
	\]

	The Kretschmann scalar of the solutions behaves as
	\[
	K=R_{αβγδ}R^{αβγδ}=(C^K+O(e^{-\nu\tau}))e^{\tau 4(p_\circ^1+p_\circ^2+p_\circ^3)}\,,
	\]
	with $C^K$ a positive constant, defined in \ref{kretsch_koeff}. As $p_\circ^1+p_\circ^2+p_\circ^3>0$ the curvature tensor grows uniformly without bounds along all causal curves.
	
	For a generic choice of the free functions, the solutions have no continuous or discrete spatial isometries, and no continuous isometries described by Killing vectors $X^μ∂_μ$ satisfying $X^τ=O(e^{-ετ})$, $ε>0$.
	
	The construction is unaffected by the presence of a cosmological constant, as demonstrated in \ref{sec:cosm_const}, giving the same asymptotic behaviour for all values of $\Lambda$.
	\appendix
	\section{Derivation of Iwasawa variable Hamiltonian}\label{sec:ham_iwa_calc}
	Here we will give the derivation of the Hamiltonian density in Iwasawa form \eqref{ham_iwa_full}, from the standard form of the Hamiltonian \eqref{ham_gij}.
	
	In the following the spatial metric and its inverse will be used in their Iwasawa forms
	\begin{equation}\label{iwa_g_ginv}
	g_{ij}=\sum_a e^{-2β^a}\N{^a_i}\N{^a_j} \quad\text{and}\quad g^{ij}=\sum_a e^{2β^a}\Ni{^i_a}\Ni{^j_a}\,.
	\end{equation}
	
	\subsection{Kinetic and symmetry wall terms}
	The kinetic term $\mathcal{K}$ and the symmetry wall term come from the first two terms in the Hamiltonian (\ref{ham_gij}). These are
	\begin{equation}\label{non_r_ham}
	π^{ij}π_{ij}-\frac{1}{2}π\indices{^i_i}π\indices{^j_j}\,.
	\end{equation}
	The conjugate momenta in Iwasawa variables, $π_a$ and $\p{^i_a}$, can be expressed in terms of $π^{ij}$ as
	\begin{equation}\label{iwa_beta_mom}
	π_a=\frac{∂\mathcal{L}}{∂\dot{β}^a}=\underbrace{\frac{∂\mathcal{L}}{∂\dot{g}_{ij}}}_{=π^{ij}}\frac{∂\dot{g}_{ij}}{∂\dot{β}^a}=-2e^{-2β^a}\N{^a_i}\N{^a_j}π^{ij}\,,
	\end{equation}
	and
	\begin{equation}\label{iwa_n_mom}
	\p{^i_a}=2π^{ij}e^{-2β^a}\N{^a_j}\,.
	\end{equation}
	We start by considering the first term in (\ref{non_r_ham}), $π^{ij}π_{ij}$. Lowering an index in the first component and raising one in the second (using the Iwasawa form (\ref{iwa_g_ginv}) of the metric) gives
	\[
	\sum_{a,b}e^{-2β^b}\N{^b_j}\N{^b_l}π\indices{_i^l}e^{2β^a}\Ni{^i_a}\Ni{^k_a}π\indices{_k^j}=\sum_{a,b}e^{2(β^a-β^b)}(π\indices{^j_k}\N{^b_j}\Ni{^k_a})^2\,.
	\]
	The double sum can be split into a diagonal and off-diagonal part
	\begin{equation}\label{diag_offdiag}
	\sum_a (π\indices{^j_k}\N{^a_j}\Ni{^k_a})^2+\sum_{a\neq b}e^{2(β^a-β^b)}(π\indices{^j_k}\N{^b_j}\Ni{^k_a})^2\,.
	\end{equation}
	Raising the index $k$ on $π\indices{^j_k}$ in the first, diagonal, part leads to
	\[
	\sum_a\left(\sum_b π^{jl}e^{-2β^b}\N{^b_k}\N{^b_l}\N{^a_j}\Ni{^k_a}\right)^2=
	\sum_a\left(e^{-2β^a}π^{jl}\N{^a_l}\N{^a_j}\right)^2=\frac{1}{4}\sum_a π_a^2\,,
	\]
	where in the last step the definition of $π_a$, (\ref{iwa_beta_mom}), was used. This, together with the second part of (\ref{non_r_ham}), gives
	{\begin{align*}
		\frac{1}{4}\sum_a π_a^2-\frac{1}{2}(g_{ij}π^{ij})^2&=\frac{1}{4}\left(\sum_a π_a^2-\frac{1}{2}(2g_{ij}π^{ij})^2\right)\\
		&=\frac{1}{4}\left(\sum_a π_a^2-\frac{1}{2}(2\sum_a e^{-2β^a}\N{^a_i}\N{^a_j}π^{ij})\right)\\
		&=\frac{1}{4}\left(\sum_a π_a^2-\frac{1}{2}(\sum_a π_a)^2\right)\\
		&=\frac{1}{4}G^{μν}π_μ π_ν\,,
		\end{align*}}
	which is the kinetic part $\mathcal{K}$ of the Hamiltonian (\ref{ham_iwa}).
	
	Raising the index $k$ in $π\indices{^j_k}$ in the second, off-diagonal, part of (\ref{diag_offdiag}) gives
	\[
	\sum_{b\neq a}e^{2(β^a-β^b)}(π^{jl}e^{-2β^a}\N{^a_l}\N{^b_j})^2
	=\sum_{b\neq a}e^{-2(β^a+β^b)}(π^{jl}\N{^a_l}\N{^b_j})^2\,.
	\]
	This is symmetric in $a$ and $b$ and can be written as
	\[
	2\sum_{a<b}e^{2(β^a-β^b)}(π^{jl}e^{-2β^a}\N{^a_l}\N{^b_j})^2=\frac{1}{2}\sum_{a<b}e^{-2(β^b-β^a)}(\p{^j_a}\N{^b_j})^2=\mathcal{V}_s\,,
	\]
	which is the potential term coming from the symmetry walls.
	
	\subsection{Gravitational wall term}
	The gravitational wall term comes from the term $-gR$ in the Hamiltonian (\ref{ham_gij}).
	
	We will calculate the curvature scalar in the Iwasawa frame. The Cartan formulas for the connection one-form $ω\indices{^a_b}$ are
	\begin{align}
	\D θ^a+\sum_b ω\indices{^a_b}\wedge θ^b=0\label{first_cartan}\,,\\
	\D γ_{ab}=ω_{ab}+ω_{ba}\label{second_cartan}\,,
	\end{align}
	where
	\[
	γ_{ab}=δ_{ab}\exp(-2β^a)=δ_{ab}A_a^2\,,
	\]
	with $A_a:=\exp(-β^a)$, is the metric in the Iwasawa frame. We will also use the definition of the structure functions
	\begin{equation}\label{wedge_structure_def}
	\Dθ^a=-\frac{1}{2}C\indices{^a_b_c}θ^b \wedge θ^c\,.
	\end{equation}
	
	$ω\indices{^a_b}$ can be obtained by considering the expression (no summation)
	\begin{equation}\label{om_start}
	A_b^2\D θ^b(e_j,e_a)+A_j^2\D θ^j(e_b,e_a)-A_a^2\Dθ^a(e_j,e_b)\,.
	\end{equation}
	Using (\ref{wedge_structure_def}) this is equal to
	\begin{equation}\label{om_left}
	-A_b^2C\indices{^b_j_a}-A_j^2C\indices{^j_b_a}+A_a^2C\indices{^a_j_b}\,.
	\end{equation}
	Starting again from (\ref{om_start}) but using (\ref{first_cartan}) gives
	\[
	A_b^2\big(ω\indices{^b_j}(e_a)-ω\indices{^b_a}(e_j)\big)+A_j^2\big(ω\indices{^j_b}(e_a)-ω\indices{^j_a}(e_b)\big)-A_a^2\big(ω\indices{^a_j}(e_b)-ω\indices{^a_b}(e_j)\big)\,.
	\]
	Lowering the upper index on $ω$ with $γ_{ab}$ and using (\ref{second_cartan}) in the form $ω_{ab}(e_c)=-ω_{ba}(e_c)+δ_{ab}(A_a^2)_{,c}$ (with $,c$ denoting the frame derivative by $e_c$) we obtain
	\begin{equation}\label{om_right}
	δ_{jb}(A_j^2)_{,a}+2ω_{ab}(e_j)-δ_{ab}(A_a^2)_{,j}-δ_{aj}(A_a^2)_{,b}\,.
	\end{equation}
	Setting (\ref{om_left}) equal to (\ref{om_right}) and raising one index using $γ^{ab}=δ^{ab}A_a^{-2}$ gives $ω\indices{^a_b}$ as
	\begin{equation}\label{om}
	ω\indices{^a_b}(e_j)=\frac{1}{2A_a^2}\left(A_b^2C\indices{^b_a_j}+A_j^2C\indices{^j_a_b}+A_a^2C\indices{^a_j_b}+δ_{ab}(A_a^2)_{,j}+δ_{aj}(A_a^2)_{,b}-δ_{jb}(A_j^2)_{,a}\right)\,.
	\end{equation}
	
	The curvature scalar can now be computed using
	\begin{equation}\label{conn_2_form}\begin{split}
	Ω\indices{^a_b}&=\D ω\indices{^a_b}+\sum_c ω\indices{^a_c} \wedge ω\indices{^c_b}\\
	&=\frac{1}{2}\sum_{e,f}R\indices{^a_b_e_f}θ^e\wedge θ^f\,,
	\end{split}\end{equation}
	as
	\begin{equation}\begin{split}
	R=\sum_{a,b,c}γ^{bc}R\indices{^a_c_a_b}&=\frac{1}{2}\sum_{e,f,a,b}A_b^{-2}R\indices{^a_b_e_f}(θ^e\wedge θ^f)(e_a,e_b)\\
	&=\sum_{a,b}A_b^{-2}Ω\indices{^a_b}(e_a,e_b)\,.
	\end{split}\end{equation}
	We start by calculating $\sum_{a,b}A_b^{-2}\D ω^a_b(e_a,e_b)$. This gives (with summation over all indices which occur more than once)
	\begin{alignat*}{2}
	&\mathrlap{A_b^{-2}\D ω^a_b(e_a,e_b)=}\\
	&\mathrlap{A_b^{-2}\left[e_a(ω\indices{^a_b}(e_b))-e_b(ω\indices{^a_b}(e_a))-C\indices{^c_a_b}ω\indices{^a_b}(e_c)\right]}\\
	&=A_b^{-2}\bigg[&&-\frac{(A_a^2)_{,a}}{2A_a^4}\left(2A_b^2C\indices{^b_a_b}+2δ_{ab}(A_b^2)_{,b}-(A_b^2)_{,a}\right)\\
	& &&+\frac{1}{2A_a^2}\left(2(A_b^2)_{,a}C\indices{^b_a_b}+2δ_{ab}(A_b^2)_{,b,b}-(A_b^2)_{,a,a}+2A_b^2C\indices{^b_a_b_{,a}}\right)\\
	& &&+\frac{(A_a^2)_{,b}}{2A_a^4}\left(2A_a^2C\indices{^a_a_b}+(A_a^2)_{,b}\right)-\frac{1}{2A_a^2}\left(2(A_a^2)_{,b}C\indices{^a_a_b}+(A_a^2)_{,b,b}+2A_a^2C\indices{^a_a_b_{,b}}\right)\\
	& &&-\frac{C\indices{^c_a_b}}{2A_a^2}\left(A_b^2C\indices{^b_a_c}+A_c^2C\indices{^c_a_b}+A_a^2C\indices{^a_c_b}+δ_{ab}(A_b^2)_{,c}+δ_{ac}(A_a^2)_{,b}δ_{cb}(A_b^2)_{,a}\right)
	\bigg]\\
	&=\mathrlap{- C\indices{^b_a_b}\frac{(A_a^2)_{,a}}{A_a^4}-\frac{(A_b^2)_{,b}(A_b^2)_{,b}}{A_b^6}+\frac{(A_b^2)_{,a}(A_a^2)_{,a}}{2A_a^4A_b^2}+2C\indices{^b_a_b}\frac{(A_b^2)_{,a}}{A_a^2A_b^2}+\frac{(A_b^2)_{,b,b}}{A_b^4}}\\
	&\phantom{{}={}}\mathrlap{-\frac{(A_b^2)_{,a,a}}{A_a^2A_b^2}+2\frac{C\indices{^b_a_b_{,a}}}{A_a^2}+\frac{(A_a^2)_{,b}(A_a^2)_{,b}}{2A_a^4A_b^2}-\frac{C\indices{^c_a_b}C\indices{^b_a_c}}{A_a^2}-\left(C\indices{^c_a_b}\right)^2\frac{A_c^2}{2A_a^2A_b^2}-C\indices{^c_b_b}\frac{(A^2)_{,c}}{2A_b^4}\,.}
	\end{alignat*}
	
	The second term, $\sum_{a,b,c}A_b^{-2}(ω\indices{^a_c}\wedge ω\indices{^c_b})(e_a,e_b)$, gives (again with all sums implied)
	{\begin{alignat*}{2}
		\mathrlap{A_b^{-2}\big(ω\indices{^a_c}(e_a)ω\indices{^c_b}(e_b)-ω\indices{^a_c}(e_b)ω\indices{^c_b}(e_a)\big)}\\
		&=\frac{1}{4A_b^2}\bigg[&&+\left(2C\indices{^a_a_c}+\frac{(A_a^2)_{,c}}{A_a^2}\right)\left(2C\indices{^b_c_b}\frac{A_b^2}{A_c^2}+\frac{2δ_{cb}(A_b^2)_{,b}}{A_b^2}-\frac{(A_b^2)_{,c}}{A_c^2}\right)\\
		& &&-\left(C\indices{^b_a_c}\frac{A_b^2}{A_a^2}+C\indices{^c_a_b}\frac{A_c^2}{A_a^2}-C\indices{^a_c_b}+\frac{δ_{ac}(A_a^2)_{,b}}{A_a^2}+\frac{δ_{ab}(A_b^2)_{,c}}{A_b^2}-\frac{δ_{cb}(A_b^2)_{,a}}{A_a^2}\right)\\
		& &&\phantom{{}-{}}\cdot\left(C\indices{^a_c_b}\frac{A_a^2}{A_c^2}+C\indices{^b_c_a}\frac{A_b^2}{A_c^2}-C\indices{^c_b_a}+\frac{δ_{cb}(A_b^2)_{,a}}{A_b^2}+\frac{δ_{ca}(A_a^2)_{,b}}{A_a^2}-\frac{δ_{ba}(A_b^2)_{,c}}{A_c^2}\right)\bigg]\\
		&\mathrlap{=C\indices{^a_a_c}C\indices{^b_c_b}\frac{1}{A_c^2}+C\indices{^a_a_b}\frac{(A_b^2)_{,b}}{A_b^4}+\frac{(A_a^2)_{,b}(A_b^2)_{,b}}{2A_a^2A_b^4}-\frac{(A_a^2)_{,c}(A_b^2)_{,c}}{4A_a^2A_b^2A_c^2}+(C\indices{^a_c_b})^2\frac{A_a^2}{4A_c^2A_b^2}}\\
		&\phantom{{}={}}\mathrlap{+C\indices{^a_c_b}C\indices{^b_c_a}\frac{1}{2A_c^2}-\frac{(A_b^2)_{,b}(A_b^2)_{,b}}{2A_b^6}+\frac{(A_b^2)_{,a}(A_b^2)_{,a}}{4A_a^2A_b^4}+C\indices{^a_c_a}\frac{(A_b^2)_{,c}}{A_c^2A_b^2}-C\indices{^b_a_b}\frac{(A_b^2)_{,a}}{A_b^2A_a^2}\,.}
		\end{alignat*}}
	Adding the two expressions and substituting $(A_a^2)_{,b}=-2A_a^2β^a_{,b}$ we obtain the curvature scalar
	\begin{equation}\begin{split}
	R=-\frac{1}{4}\sum_{a,b,c}(C\indices{^a_c_b})^2\frac{A_a^2}{A_c^2A_b^2}+\sum_a\bigg\{-2\frac{(β^a_{,a})^2}{A_a^2}-2\frac{β^a_{,a,a}}{A_a^2}+\sum_b\bigg[-4C\indices{^a_a_b}\frac{β^b_{,b}}{A_b^2}+4\frac{β^a_{,b}β^b_{,b}}{A_b^2}\\
	-2C\indices{^b_a_b}\frac{β^b_{,a}}{A_a^2}-\frac{(β^b_{,a})^2}{A_a^2}+2\frac{β^b_{,a,a}}{A_a^2}+2\frac{C\indices{^b_a_b_{,a}}}{A_a^2}+C\indices{^a_b_b}\frac{β^b_{,a}}{A^2_b}\\
	+\sum_c\bigg(\frac{1}{A_c^2}C\indices{^a_a_c}C\indices{^b_c_b}-\frac{β^a_{,c}β^b_{,c}}{A_c^2}-\frac{1}{2A_c^2}C\indices{^a_c_b}C\indices{^b_c_a}-2C\indices{^a_c_a}\frac{β^b_{,c}}{A_c^2}\bigg)\bigg]\bigg\}\,.
	\end{split}\end{equation}
	
	Multiplying this with $-g=-\exp(-2\sum_a β^a)$ gives the gravitational wall terms (\ref{dom_grav_walls}) and (\ref{subdom_grav_walls}) in the Iwasawa variable Hamiltonian.

	\section{Iwasawa evolution equations and Einstein equations}\label{sec:equiv_eeq}
	
	To obtain the evolution equations \eqref{evo_eqs} the variation was taken after choosing lapse $N$ and shift $N_a$ with the lapse given as $\sqrt{\det g}$, \ie dependent on the metric, and the shift vanishing.
	
	The general Hamiltonian density, with lapse and shift still free, is
	\begin{equation}\label{wald_ham}\begin{split}
	H=\sqrt{\det g}\bigg\{&N\left[-R+(\det g)^{-1}π^{ij}π_{ij}-\frac{1}{2}(\det g)^{-1}(π^i_i)^2\right]\\&-2N_j[\nabla_i((\det g)^{-1/2}π^{ij})]+2\nabla_i((\det g)^{-1/2}N_jπ^{ij})\bigg\}\,,
	\end{split}\end{equation}
	(Equation (E.2.32) in Wald \cite{Wald1984}).
	
	Taking the variation of the Hamiltonian $\bar{H}=\int H \D^3x$ with regards to $N$ and $N_a$ gives the Hamiltonian and momentum constraints, respectively:
	\begin{align}
	\label{wald_ham_const_app}\frac{δ\bar{H}}{δN}&=-R+(\det g)^{-1}π^{ij}π_{ij}-\frac{1}{2}(\det g)^{-1}(π^i_i)^2\,,\\
	\label{wald_mom_const_app}\frac{δ\bar{H}}{δN_i}&=\nabla_j((\det g)^{-1/2} π^{ji})=\nabla_j π^{ji}\,.
	\end{align}
	
	Varying now with respect to $π^{ij}$ and $g_{ij}$ gives the Einstein equations in Hamiltonian form as
	\begin{alignat}{2}\label{wald_evo_g_eq}
	\dot{g}_{ij}&=\mathrlap{\frac{δ\bar{H}}{δπ^{ij}}=2(\det g)^{-1/2}N\left(π_{ij}-\frac{1}{2}g_{ij}π^k_k\right)+2\nabla_{(i}N_{j)}\,,}\\
	\label{wald_evo_pi_eq}\dot{π}^{ij}&=-\frac{δ\bar{H}}{δg_{ij}}=&&-N\sqrt{\det g}\left(R^{ij}-\frac{1}{2}Ng^{ij}\right)+\frac{1}{2}N(\det g)^{-1/2}g^{ij}\left(π_{kl}π^{kl}-\frac{1}{2}(π^k_k)^2\right)\\
	& &&\nonumber-2N(\det g)^{-1/2}\left(π^a{}_c π\indices{_c^b}-\frac{1}{2}π^k{}_kπ^{ij}\right)-\sqrt{\det g}(\nabla^i\nabla^j N-g^{ij}\nabla^k\nabla_k N)\\
	& &&\nonumber+\nabla_k(N^kπ^{ij})-2π^{k(i}\nabla_cN^{j)}\,,
	\end{alignat}
	(Equations (E.2.35) and (E.2.36) in Wald \cite{Wald1984}).
	
	Choosing $N_a=0$, either before or after varying, just removes the terms containing $N_a$ in the evolution equations. Choosing $N=\sqrt{\det g}$ before varying adds an additional term in \eqref{wald_evo_pi_eq}. This term is, however, proportional to $-R+(\det g)^{-1}π^{ij}π_{ij}-\frac{1}{2}(\det g)^{-1}(π^i_i)^2$ which is zero, by the Hamiltonian constraint \eqref{wald_ham_const_app}.
	
	The terms in \eqref{wald_evo_pi_eq} which contain covariant derivatives of the lapse also vanish, as the determinant of the metric is covariantly constant.
	
	The transformation to Iwasawa variables is a point canonical transformation and therefore doesn't change the equations.

	\section{Derivation of Iwasawa variable momentum constraints}\label{sec:calc_iwa_mom_const}
	In this section we will give the derivation of the momentum constraints in Iwasawa variables and the definition of their asymptotic equivalent, following section 3.2 of \cite{Damour2008}.
	\subsection{Full momentum constraints}\label{sec_full_mom_const_deriv}
	We start with the momentum constraints in the form
	\begin{equation}\label{wald_mom_const}
	\nabla_i π\indices{^i_j}=0
	\end{equation}
	(see e.g. equation (E.2.34) in Wald \cite{Wald1984}).
	
	The calculation is simpler when done in the Iwasawa frame (\ref{iwa_frame}), so we first calculate the Iwasawa frame components of $π^{ij}$, in terms of the Iwasawa variable conjugate momenta $P\indices{^i_a}$ and $π_a$. These are denoted by $\tilde{π}^{ab}$ and defined as
	\[
	\tilde{π}\indices{^a^b}:=π(θ^a,θ^b)=π^{ij}(θ^a)_i(θ^b)_j=π^{ij}\N{^a_i}\N{^b_j}\,.
	\]
	
	Starting from $\dot{g}_{ij}π^{ij}=\dot{β}^aπ_a+\dot{\mathcal{N}}\indices{^a_i}\p{^i_a}$ and writing the left side in Iwasawa variables gives
	\begin{equation}
	\sum_a 2 e^{-2β^a}\left(\dot{\mathcal{N}}\indices{^a_i}\Ni{^i_c}\tilde{π}\indices{^c^a}-\dot{β}^a \tilde{π}\indices{^a^a}\right)=\dot{β}^aπ_a+\dot{\mathcal{N}}\indices{^a_i}\p{^i_a}\,.
	\end{equation}
	Using the diagonal form of the metric in the Iwasawa frame, $γ_{ab}=\exp(-2β^a)δ_{ab}$, we obtain
	\begin{alignat}{2}
	\tilde{π}\indices{^a_a}&=-\frac{1}{2}π_a&\qquad\text{(no summation),}\label{pi_iwa_diag}\\
	\Ni{^i_c}\tilde{π}\indices{^c_a}&=\frac{1}{2}\p{^i_a}&\qquad\text{for }i>a\,,
	\end{alignat}
	by comparison. This can be written as a matrix equation
	\begin{equation}\label{pi_iwa_matrixeq}
	\Ni{^i_c_{\,(+)}}\tilde{π}\indices{^c_a}=\frac{1}{2}\p{^i_a_{\,[-]}}+X\indices{^i_a_{\,(+)}}
	\end{equation}
	(for all $i$ and $a$) where the subscript $(+)/(-)$ designates an upper/lower triangular matrix and $[+]/[-]$ a strictly upper/lower triangular one. The matrix $X\indices{^i_a_{(+)}}$ is defined by this equation.
	
	Multiplying (\ref{pi_iwa_matrixeq}) by $\N{^b_i}=\N{^b_i_{\,(+)}}$ gives
	\[
	\tilde{π}\indices{^b_a}=\frac{1}{2}\N{^b_i_{\,(+)}}\p{^i_a_{\,[-]}}+\N{^b_i_{\,(+)}}X\indices{^i_a_{(+)}}\,.
	\]
	The strictly lower triangular part of $\tilde{π}\indices{^b_a}$, $\tilde{π}\indices{^b_a_{\,[-]}}$, is given explicitly by
	\[
	\tilde{π}\indices{^b_a_{\,[-]}}=\frac{1}{2}\N{^b_i_{\,(+)}}\p{^i_a_{\,[-]}}θ(b-a)\quad\text{with}\quad θ(x):=
	\begin{cases}
	0&\text{if }x\leq 0\,,\\
	1&\text{if }x>0\,.
	\end{cases}
	\]
	Because of the symmetry of $\tilde{π}^{ab}$ this also gives the upper triangular part via
	\begin{align*}
	\tilde{π}\indices{^b_a_{\,[+]}}=&θ(a-b)e^{-2β^a}\tilde{π}^{ba}=θ(a-b)e^{-2β^a}\tilde{π}^{ab}\\
	=&e^{-2(β^a-β^b)}\tilde{π}\indices{^a_b}θ(a-b)=e^{-2(β^a-β^b)}\tilde{π}\indices{^a_b_{\,[-]}}\,.
	\end{align*}
	Finally, also including the diagonal term from (\ref{pi_iwa_diag}), we arrive at
	\begin{equation}\label{pi_iwa}
	\tilde{π}\indices{^b_a}=\frac{1}{2}\begin{cases}
	-π_b&\text{for }a=b\,,\\
	\N{^b_i}\p{^i_a}&\text{for }b>a\,,\\
	e^{-2(β^a-β^b)}\N{^a_i}\p{^i_b}&\text{for }a>b\,.
	\end{cases}
	\end{equation}
	
	Writing the momentum constraint (\ref{wald_mom_const}) in the Iwasawa frame and expanding the covariant derivative gives
	\begin{equation}\label{mom_const_iwa_frame}
	\nabla_b\tilde{π}\indices{^b_a}=\tilde{π}\indices{^b_a_{,b}}+Γ\indices{^b_d_b}\tilde{π}\indices{^d_a}-Γ\indices{^d_a_b}\tilde{π}\indices{^b_d}-Γ\indices{^c_c_b}\tilde{π}\indices{^b_a}\,,
	\end{equation}
	where the last term comes from the fact that $\tilde{π}\indices{^b_a}$ is a tensor density of weight 1 and $Γ\indices{^a_b_c}$ are the connection coefficients in the Iwasawa frame, given by
	\begin{equation}\label{iwa_frame_conn_coeff}\begin{alignedat}{2}
	Γ\indices{^a_b_c}&=\mathrlap{\frac{1}{2}\sum_σ g^{aσ}(g_{σc,b}+g_{bσ,c}-g_{bc,σ}-C_{σbc}+C_{bσc}+C_{cσb})}\\
	&=\frac{1}{2}e^{2β^a}\big(&&δ_{ac}(e^{-2β^a})_{,b}+δ_{ab}(e^{-2β^a})_{,c}-δ_{bc}(e^{-2β^b})_{,a}\\
	& &&-e^{-2β^a}C\indices{^a_b_c}+e^{-2β^b}C\indices{^b_a_c}+e^{-2β^c}C\indices{^c_a_b}\big)\,,
	\end{alignedat}\end{equation}
	(no implicit summation) with the comma denoting the derivative in the Iwasawa frame.
	
	Inserting (\ref{iwa_frame_conn_coeff}) into (\ref{mom_const_iwa_frame}) gives
	\begin{equation}\label{full_mom_const_derived}
	\nabla_b\tilde{π}\indices{^b_a}=\tilde{π}\indices{^b_a_{,b}}+C\indices{^c_c_b}\tilde{π}\indices{^b_a}+C\indices{^b_a_c}\tilde{π}\indices{^c_b}-\frac{1}{2}β\indices{^b_{,a}}π_b\,,
	\end{equation}
	which is the momentum constraint (\ref{full_mom_const}).
	
	\subsection{Asymptotic momentum constraints}
	The asymptotic momentum constraints are obtained from the full ones by discarding the contribution of the strictly upper triangular part of $\tilde{π}\indices{^b_a}$: Indeed, this part (the last line of (\ref{pi_iwa})) contains exponential terms which vanish asymptotically if the symmetry wall conditions are fulfilled.
	
	Splitting $\tilde{π}\indices{^b_a}$ into $\tilde{π}\indices{^b_a}=\tilde{π}\indices{^b_a_{\,[+]}}+\tilde{π}\indices{^b_a_{\,[-]}}-δ^a_bπ_a/2$, discarding $\tilde{π}\indices{^b_a_{\,[+]}}$ and inserting into (\ref{full_mom_const_derived}) gives
	\begin{align*}
	\displaystyle\nabla_b\tilde{π}\indices{^b_a}
	\xrightarrow{\scriptscriptstyle τ\to\infty}\;&
	\tilde{π}\indices{^b_a_{[-]}_{,b}}-\frac{1}{2}π_{a,a}+C\indices{^c_c_b}\tilde{π}\indices{^b_a_{[-]}}+C\indices{^b_a_c}\tilde{π}\indices{^c_b_{[-]}}\\
	&-\frac{1}{2}C\indices{^c_c_a}π_a-\frac{1}{2}C\indices{^b_a_b}π_b
	-\frac{1}{2}β\indices{^b_{,a}}π_b
	\,,
	\end{align*}
	which is the asymptotic momentum constraint (\ref{asym_mom_const}).
	
	Inserting the ansatz (\ref{myansatz}) (which implies $C\indices{^a_b_c}=0$ asymptotically) and the asymptotic evolution of the $β^a$, $β_{[0]}^a=p_\circ^a τ+β_\circ^a$ gives
	\[\begin{split}
	\nabla_b\tilde{π}\indices{^b_a}\xrightarrow{\scriptscriptstyle τ\to\infty}\;&\sum_b\tilde{π}\indices{_{[0]}^b_a_{[-]}_{,b}}-\frac{1}{2}π\indices{_{[0]}_a_{,a}}-\frac{1}{2}\sum_b β\indices{_{[0]}^b_{,a}}π_{[0]b}\\
	&=\sum_{\substack{b>a\\i}}\frac{1}{2}\N{_\circ^b_i}\p{_\circ^i_a_{,b}}-\sum_b G_{ab}p_{\circ,a}^b-\sum_{b,c}(β^b_{\circ,a}+τp_{\circ,a}^b)G_{bc}p_{\circ}^c\,.
	\end{split}\]
	The last term, which contains a time dependence, is zero if the asymptotic Hamiltonian constraint $G_{bc}p_\circ^bp_\circ^c=0$ is fulfilled. Expressing the Iwasawa frame derivative in terms of the coordinate derivative $∂_i$ leads to
	\[
	\nabla_b\tilde{π}\indices{^b_a}\xrightarrow{\scriptscriptstyle τ\to\infty}\;\frac{1}{2}\sum_{\substack{b>a\\i,j}}\N{_\circ^b_i}(\N{_\circ^{-1}})\indices{^j_b}∂_jP\indices{_\circ^i_a}-\sum_{j}(\N{_\circ^{-1}})\indices{^j_a}\left(\sum_c G_{ac}∂_jp_{\circ}^c + \sum_{d,f} G_{df}p_\circ^f ∂_jβ^d_{\circ}\right)\,,
	\]
	which is (\ref{my_asym_mom_const}) with arbitrary (constant) $\N{_\circ^a_i}$.
	
	\section{Evolution equations for the constraints}\label{sec:const_evo}
	Here we will give the derivation of the evolution equations for the constraints in our choice of gauge. Our treatment is similar to, but not identical with  appendix A of \cite{Damour2008}.
	
	The first order action corresponding to the Einstein equations is given by
	\begin{equation}\label{17III15.1}
	S[g_{ij},π^{ij},\tilde{N},N^i]=\int \dx^0\D^d x(\dot{g}_{ij}π^{ij}-\tilde{N}H-N^iH_i)\,,
	\end{equation}
	with $\tilde{N}$ the ``rescaled lapse'' defined as $\tilde{N}=N/\sqrt{g}$ and $N^i$ the shift vector. In our choice of gauge $\tilde{N}=1$ and $N^i=0$. From \eqref{17III15.1} the equations of motions, the Hamiltonian constraints and the momentum constraints can be obtained by varying with respect to $g_{ij}$, $\tilde{N}$ and $N^i$ respectively.
	
	We will compare the resulting equations with those coming from variation of the standard Einstein-Hilbert action $S_H=\int \D^D x\sqrt{-\bar{g}}\bar{R}$, which is (neglecting boundary terms)
	\begin{equation}\label{17III15.2}
	δS_H=-\int \D^Dx \sqrt{-\bar{g}}\bar{G}^{μν}δ\bar{g}_{μν}\,,
	\end{equation}
	with $\bar{G}$ denoting the Einstein tensor.
	
	As the spacetime metric $\bar{g}_{μν}$ is defined, in terms of $\tilde{N}$, $N_i$ and $g_{ij}$, as
	\[
	(\bar{g}_{μν})=\begin{pmatrix}
	N_kN^k-\tilde{N}^2 g&N_j\\
	N_i&g_{ij}
	\end{pmatrix}\,,
	\]
	the variation of $S_H$ with respect to $g_{ij}$, $\tilde{N}$ and $N_i$ following from \eqref{17III15.2} is given by
	\begin{align}
	\label{17III15.3}
	\frac{δS_H}{δg_{ij}}&=\tilde{N}g(-\bar{G}^{ij}+\tilde{N}^2gG^{00}g^{ij})+O(N^k)\,,\\
	\label{17III15.4}
	\frac{δS_H}{δ\tilde{N}}&=2\bar{G}^{00}\tilde{N}^2g^2+O(N^k)\,,\\
	\label{17III15.5}
	\frac{δS_H}{δN_i}&=-2\tilde{N}g\bar{G}^{i0}+O(N^k)\,,
	\end{align}
	where $O(N^k)$ denotes terms proportional to $N^k$ which vanish in our gauge. Here $\sqrt{-\bar{g}}=\tilde{N}g$ and  $δg=gg^{ij}δg_{ij}$ were used.
	
	From the first order action \eqref{17III15.1} we obtain
	\begin{align}
	\label{17III15.6}
	\frac{δS}{δ\tilde{N}}&=-H\,,\\
	\label{17III15.7}
	\frac{δS}{δN_i}&=-H^i\,.
	\end{align}
	Identifying \eqref{17III15.4} with \eqref{17III15.6} and \eqref{17III15.5} with \eqref{17III15.7} yields
	\begin{align}
	\label{17III15.8}
	H&=-2g^2\tilde{N}^2\bar{G}^{00}+O(N^k)=-\frac{2}{\tilde{N}^2}\bar{G}_{00}+O(N^k)\,,\\
	\label{17III15.9}
	H^i&=2g\tilde{N}\bar{G}^{i0}+O(N^k)=-\frac{2}{\tilde{N}}g^{ij}\bar{G}_{0j}+O(N^k)\,.
	\end{align}
	Using \eqref{17III15.8} to rewrite the equations of motion $δS_H/δg_{ij}=0$ from \eqref{17III15.3} leads to
	\begin{equation}\label{17III15.10}
	\bar{G}^{ij}=-\frac{1}{2g}Hg^{ij}+O(N^k)\,.
	\end{equation}
	
	We now consider the (vanishing) divergence of the Einstein tensor $\nabla_ν\bar{G}^ν{}_μ=0$. Using the identity
	\[
	Γ^ν{}_{νμ}=\frac{∂_μ\sqrt{-\bar{g}}}{\sqrt{-\bar{g}}}\,,
	\]
	this can be rewritten as
	\begin{align*}
	0=\nabla_ν\bar{G}^ν{}_μ&=∂_ν\bar{G}^ν{}_μ+Γ^ν{}_{να}\bar{G}^α{}_μ-Γ^α{}_{νμ}\bar{G}^ν{}_α\\
	&=∂_ν\bar{G}^ν{}_μ+\frac{∂_α\sqrt{-\bar{g}}}{\sqrt{-\bar{g}}}\bar{G}^α{}_μ-\frac{1}{2}g^{ασ}(∂_νg_{σμ}+∂_μg_{νσ}-∂_σg_{νμ})\bar{G}^ν{}_μ\\
	&=\frac{∂_ν(\bar{G}^ν{}_μ\sqrt{-\bar{g}})}{\sqrt{-\bar{g}}}-\frac{1}{2}\bar{G}^{νσ}∂_μg_{νσ}\,.
	\end{align*}
	Expressing the components of the Einstein tensor using \eqref{17III15.8}, \eqref{17III15.9} and \eqref{17III15.10} gives for $μ=0$ (with $∂_τ=∂_0$)
	\begin{equation}\label{17III15.11}\begin{aligned}
	O(N^k)&=\frac{1}{\tilde{N}g}\left(∂_i(g_{00}H^i/2)-∂_τ\left(g_{00}\frac{H}{2g\tilde{N}}\right)\right)-\frac{H}{4g^2\tilde{N}^2}∂_τ(\tilde{N}^2g)+\frac{H}{4g}g^{ij}∂_τg_{ij}\\
	&=-\frac{H^i}{2\tilde{N}g}∂_i(\tilde{N}g)-\frac{\tilde{N}}{2}∂_iH^i+\frac{1}{2g}∂_τH+\frac{H}{2\tilde{N}g}∂_τ\tilde{N}-\frac{H}{4g^2\tilde{N}^2}∂_τ(\tilde{N}^2g)+\frac{H}{4g}g^{ij}∂_τg_{ij}\,.
	\end{aligned}\end{equation}
	The last three terms cancel as $∂_τg=gg^{ij}∂_τg_{ij}$.
	
	Doing the same for the $μ=i$ equations, we obtain
	\begin{equation}\label{18III15.1}
	O(N^k)=\frac{∂_τH_i}{2\tilde{N}g}-\frac{1}{2}∂_i\left(\frac{H}{g}\right)-\frac{H}{2g^2}∂_ig-\frac{H}{\tilde{N}}∂_i\tilde{N}\,.
	\end{equation}
	
	As $\tilde{N}$ is a scalar density of weight $-1$, $H$ a scalar density of weight $2$ and $H_i$ a tensor density of weight $1$ their covariant derivatives are given by
	\[
	\nabla_i\tilde{N}=∂_i(\tilde{N}\sqrt{g})/\sqrt{g}\,,\quad
	\nabla_i H=g∂_i(H/g)\quad\text{and}\quad
	\nabla_i H_j=∂_iH_j-Γ^k{}_{ij}H_k-Γ^k{}_{ki}H_j\,.
	\]
	The divergence of $H^i$ can therefore be expressed as $\nabla_iH^i=∂_i(g^{ij}H_i)$\,. Inserting this into \eqref{17III15.11} and \eqref{18III15.1} and applying our gauge choice $N^k=0$, $\tilde{N}=1$ yields
	\begin{align}
	\label{23III15.1}
	∂_τH&=g\nabla_i H^i+H^i∂_ig\,,\\
	\label{23III15.2}
	∂_τH_i&=\nabla_i H+\frac{H}{g}∂_ig\,.
	\end{align}
	The last term of equation \eqref{23III15.2} seems to have been overlooked in the derivation of \cite{Damour2008}. It does not affect the arguments of section \ref{sec:asym_full_const} concerning the relationship between asymptotic and full constraints.
	
	Equations \eqref{23III15.1} and \eqref{23III15.2} can be expressed in the Iwasawa frame as
	\begin{align}
	∂_τH=g\nabla_a H^a+H^a\tilde{∂}_a g\,,\\
	∂_τH_a=\nabla_a H+\frac{H}{g}\tilde{∂}_ag\,,
	\end{align}
	using $H_i\dx^i=H_aθ^a=H_a\N{^a_i}\dx^i$ with $H_a$ the components of the momentum constraint in the Iwasawa frame and $\tilde{∂}_a=(\N{^{-1}})\indices{^i_a}∂_i$ the Iwasawa frame derivative.

	\section{Asymptotic behaviour of the metric and curvature}\label{sec:asym_beh}
	The ansatz \eqref{myansatz} and the Fuchs theorem give for the asymptotic behaviour of the matrix $\N{}$
	\[
	\N{^a_i}=δ^a_i+e^{-γτ}\N{_s^a_i}=δ^a_i+O(e^{-τ(γ+ν)})\,.
	\]
	
	An explicit first order term for the off-diagonal elements can be found from the evolution equation for $ \N{_s^a_i}$, \eqref{my_n_eq}. This equation is
	\begin{equation}\label{16VI15.1}
	∂_τ\N{_s^a_i}=γ\N{_s^a_i}+e^{γτ}\sum_A\frac{∂c_A}{∂P\indices{^i_a}}e^{-2w_A(β_{[0]})}e^{-2w_A(\bar{β})}\,.
	\end{equation}
	The sum over the potential walls includes only the symmetry walls, as the coefficients of the others do not depend on $\p{^i_a}$. Their coefficients are $c_{ab}=(\p{^j_a}\N{^b_j})^2/2$, which gives, inserted into \eqref{16VI15.1},
	\begin{equation}\label{16VI15.2}
	∂_τ\N{_s^a_i}=γ\N{_s^a_i}+e^{γτ}\sum_{c>a}(\p{^j_a}\N{^c_j}\N{^c_i})e^{-2(β^c-β^a)}\,.
	\end{equation}
	Inserting the asymptotic behaviour of $\p{}$ and $\N{}$ gives
	\begin{equation}\label{16VI15.3}
	∂_τ\N{_s^a_i}=γ\N{_s^a_i}+e^{γτ}\sum_{ c>a}\big((\underbrace{\p{_\circ^j_a}+\bar{P}\indices{^j_a}}_{A})(\underbrace{δ^c_j+e^{-γτ}\N{_s^c_j}}_{B})(\underbrace{δ^c_i+e^{-γτ}\N{_s^c_i}}_{C})\big)e^{-2(β^c-β^a)}\,.
	\end{equation}
	We consider first term $A$. As $\p{_\circ^j_a}$ is of order $1$ while $\bar{P}\indices{^j_a}$ decays as $e^{-ντ}$ by the Fuchs theorem, the first term dominates. For term $B$ the situation is similar. The $δ^c_j$ term is of order one for $c=j$ (the other terms in the $j$-sum are irrelevant as they share the same decay rate) while the second term decays as $e^{-τ(γ+ν)}$.
	
	This leaves
	\begin{equation}
	∂_τ\N{_s^a_i}=γ\N{_s^a_i}+e^{γτ}\sum_{c, c>a}\big((\p{_\circ^c_a}+O(e^{-ντ}))(\underbrace{δ^c_i+e^{-γτ}\N{_s^c_i}}_{C})\big)e^{-2(β^c-β^a)}\,.
	\end{equation}
	
	For $C$ the situation is different, as the sum over $c$ also influences the factor $\exp(-2(β^c-β^a))$. We will consider the three components of $\N{_s^a_i}$ separately. In the case $\N{_s^1_2}$, \ie $a=1$ and $i=2$, only the summand with $c=2$ survives in the sum, as $\N{_s^c_i}=\N{_s^c_2}$ is zero for $c=2$ and $c=3$ because of the upper triangular form of $\N{_s}$. The evolution equation therefore becomes
	\[
	∂_τ\N{_s^1_2}=γ\N{_s^1_2}+e^{γτ}\p{_\circ^2_1}e^{-2(β^2-β^1)}+O(e^{-τ(2p_\circ^2-2p_\circ^1-γ+ν)})\,.
	\]
	Integrating this and using the fact that $\N{_s^1_2}=O(e^{-ντ})$ from the Fuchs theorem to eliminate the integration constant gives
	\begin{equation}\label{16VI15.4}
	\N{_s^1_2}=\underbrace{-\frac{\p{_\circ^2_1}e^{-2(β_\circ^2-\beta_\circ^1)}}{2(p_\circ^2-p_\circ^1)}}_{=:K\indices{^1_2}}e^{-τ(2p_\circ^2-2p_\circ^1-γ)}+O(e^{-τ(2p_\circ^2-2p_\circ^1-γ+ν)})\,.
	\end{equation}
	For $\N{_s^2_3}$ the situation is similar: $\N{_s^3_3}$ vanishes, leaving only the $δ^c_i$ term and giving
	\[
	∂_τ\N{_s^2_3}=γ\N{_s^2_3}+e^{γτ}\p{_\circ^3_2}e^{-2(β^3-β^2)}+O(e^{-τ(2p_\circ^3-2p_\circ^2-γ+ν)})\,,
	\]
	and therefore
	\begin{equation}\label{16VI15.5}
	\N{_s^2_3}=\underbrace{-\frac{\p{_\circ^3_2}e^{-2(β_\circ^3-β_\circ^2)}}{2(p_\circ^3-p_\circ^2)}}_{=:K\indices{^2_3}}e^{-τ(2p_\circ^3-2p_\circ^2-γ)}+O(e^{-τ(2p_\circ^3-2p_\circ^2-γ+ν)})\,.
	\end{equation}
	For $\N{_s^1_3}$ two summands remain: For $c=2$ the term with $\N{_s^2_3}$ and for $c=3$ the one with $δ^c_i$. The evolution equation becomes
	\[
	∂_τ\N{_s^1_3}=γ\N{_s^1_3}+\p{_\circ^2_1}\N{_s^2_3}e^{-2(β^2-β^1)}+e^{γτ}\p{_\circ^3_1}e^{-2(β^3-β^1)}\,.
	\]
	Inserting $\N{_s^2_3}$ from \eqref{16VI15.5} gives
	\[
	∂_τ\N{_s^1_3}=γ\N{_s^1_3}+e^{-2(β_\circ^3-β_\circ^1)}\bigg(\p{_\circ^3_1}-\frac{\p{_\circ^2_1}\p{_\circ^3_2}}{2p_\circ^3-2p_\circ^2}\bigg)e^{-τ(2p_\circ^3-2p_\circ^1-γ)}+O(e^{-τ(2p_\circ^3-2p_\circ^1-γ+ν)})
	\]
	and after integrating
	\begin{equation}\label{16VI15.6}
	\N{_s^1_3}=\underbrace{e^{-2(β_\circ^3-β_\circ^1)}\bigg(\p{_\circ^3_1}-\frac{\p{_\circ^2_1}\p{_\circ^3_2}}{2p_\circ^3-2p_\circ^2}\bigg)\frac{1}{2p_\circ^3-2p_\circ^1}}_{=:K\indices{^1_3}}e^{-τ(2p_\circ^3-2p_\circ^1-γ)}+O(e^{-τ(2p_\circ^3-2p_\circ^1-γ+ν)})\,.
	\end{equation}
	
	From equations \eqref{16VI15.4}, \eqref{16VI15.5} and \eqref{16VI15.6} the full matrix $\N{^a_i}$ and its inverse $\Ni{^i_a}$ are given by
	\begin{align*}
	\N{^1_2}&=(K\indices{^1_2}+O^ν)e^{-τ(2p_\circ^2-2p_\circ^1)} \,,
	&\Ni{^1_2}&=-(K\indices{^1_2}+O^ν)e^{-τ(2p_\circ^2-2p_\circ^1)} \,,\\
	\N{^1_3}&=(K\indices{^1_3}+O^ν)e^{-τ(2p_\circ^3-2p_\circ^1)} \,,
	&\Ni{^1_3}&=-(K\indices{^1_3}+O^ν)e^{-τ(2p_\circ^3-2p_\circ^1)} \,,\\
	\N{^2_3}&=(K\indices{^2_3}+O^ν)e^{-τ(2p_\circ^3-2p_\circ^2)} \,,
	&\Ni{^2_3}&=-(K\indices{^2_3}+O^ν)e^{-τ(2p_\circ^3-2p_\circ^2)} \,,
	\end{align*}
	where $O^ν:=O(e^{-ντ})$.

	Setting
	\begin{equation}
	\sigmap = p^1_{\circ} + p^2_{\circ}  + p^3_{\circ}  >0
	\;,
	\quad
	\sigmab = \beta^1_{\circ} + \beta^2_{\circ}  + \beta^3_{\circ}
	\;,
	\end{equation}
	the leading order behaviour of the metric components, and those of its inverse, is
	\begin{equation}\label{2015VII6.2}
	\begin{aligned}
	\bar{g}_{00} & =  -e^{-2\sigmap τ - 2 \sigmab }(1 + O^ν)\to0
	\,,&
	\bar{g}^{00}&=-e^{2\sigmap τ + 2 \sigmab }(1 + O^ν)
	\,,
	\\
	\bar{g}_{0 i} & \equiv  0
	\,,&
	\bar{g}^{0 i} & \equiv 0
	\,,
	\\
	\bar{g}_{11} & = e^{-2p^1_{\circ} \tau -2  \beta^1_{\circ}  }(1 + O^ν)
	\,,&
	\bar{g}^{11} & = e^{2p^1_\circ τ + 2 β^1_\circ}(1 + O^ν)\to0
	\,,
	\\
	\bar{g}_{22} & = e^{-2p^2_{\circ} \tau -2  \beta^2_{\circ}  }(1 + O^ν)\to0
	\,,&
	\bar{g}^{22} & = e^{2p^2_\circ τ + 2 β^2_\circ}(1 + O^ν)
	\,,
	\\
	\bar{g}_{33} & = e^{-2p^3_{\circ} \tau -2  \beta^3_{\circ}  }(1 + O^ν)\to0
	\,,&
	\bar{g}^{33} & = e^{2p^3_\circ τ + 2 β^3_\circ}(1 + O^ν)
	\,,
	\\
	\bar{g}_{12} & = e^{-2p_\circ^2τ}e^{-2β_\circ^1}(K\indices{^1_2}+O^ν)\to0
	\,,&
	\bar{g}^{12} & = -e^{2p_\circ^1τ}e^{2β_\circ^2}(K\indices{^1_2}+O^ν)\to 0
	\,,
	\\
	\bar{g}_{13} & = e^{-2p_\circ^3τ}e^{-2β_\circ^1}(K\indices{^1_3}+O^ν)\to 0
	\,,&
	\bar{g}^{13} & = -e^{2p_\circ^1τ}e^{2β_\circ^3}(K\indices{^1_3}+O^ν)\to 0
	\,,
	\\
	\bar{g}_{23} & = e^{-2p_\circ^3τ}e^{-2β_\circ^2}(K\indices{^2_3}+O^ν)\to 0
	\,,&
	\bar{g}^{23} & = -e^{2p_\circ^2τ}e^{2β_\circ^3}(K{^2_3}+O^ν)
	\,,
	\end{aligned}
	\end{equation}
	where all decaying ones have been marked with ``$\to0$''.
	
	The Christoffel symbols are defined as
	\begin{equation}\label{2015VII6.1}
	Γ\indices{^\alpha_\beta_\gamma}=\frac{1}{2} \bar{g}^{ασ} (\bar{g}_{σγ,β}+\bar{g}_{βσ,γ}-\bar{g}_{βγ,σ})\,,
	\end{equation}
	and show the following behaviour:
	{\allowdisplaybreaks\begin{align*}
		\Gamma\indices{^0_0_0}&=-\sigmap(1+O(e^{-ντ}))\,, &
		Γ\indices{^0_i_0}&=(τ σ_{p_\circ,i}+σ_{β_\circ,i})(1+O(e^{-ντ}))\,,&\\
		Γ\indices{^i_0_i}&=-p_\circ^i(1+O(e^{-ντ}))\quad\text{(no sum)}\,,&
		\Gamma\indices{^0_1_1}&=-p_\circ^1 e^{2(β_\circ^2+β_\circ^3)}e^{2(p_\circ^2+p_\circ^3)τ}(1+O(e^{-ντ}))\,, &\\
		\Gamma\indices{^0_2_2}&=-p_\circ^2 e^{2(β_\circ^1+β_\circ^3)}e^{2(p_\circ^1+p_\circ^3)τ}(1+O(e^{-ντ}))\,, &
		\Gamma\indices{^0_3_3}&=-p_\circ^3 e^{2(β_\circ^1+β_\circ^2)} e^{2(p_\circ^1+p_\circ^2)τ}(1+O(e^{-ντ}))\,, &\\
		\Gamma\indices{^0_2_1}&=(C+O^ν)e^{2(p_\circ^1+p_\circ^3)\tau}\,, &
		\Gamma\indices{^0_3_1}&=(C+O^ν)e^{2(p_\circ^1+p_\circ^2)\tau}\,, &\\
		\Gamma\indices{^0_3_2}&=(C+O^ν)e^{2(p_\circ^1+p_\circ^2)\tau}\,, &
		\Gamma\indices{^1_0_0}&=(C+O^ν)τe^{-2(p_\circ^2+p_\circ^3)\tau}\to 0\,, &\\
		\Gamma\indices{^1_1_1}&=(C+O^ν)τ\,, &
		\Gamma\indices{^1_2_0}&=o(e^{2(p_\circ^1-p_\circ^3)})\to 0\,, &\\
		\Gamma\indices{^1_2_1}&=(C+O^ν)τ\,, &
		\Gamma\indices{^1_2_2}&=(C+O^ν)τ e^{2(p_\circ^1-p_\circ^2)\tau}\to 0\,, &\\
		\Gamma\indices{^1_3_0}&=(C+O^ν)e^{2(p_\circ^1-p_\circ^3)\tau}\to 0\,, &
		\Gamma\indices{^1_3_1}&=(C+O^ν)τ\,, &\\
		\Gamma\indices{^1_3_2}&=o(τe^{2(p_\circ^1-p_\circ^2)τ})\to 0\,, &
		\Gamma\indices{^1_3_3}&=(C+O^ν)τe^{2(p_\circ^1-p_\circ^3)\tau}\to 0\,, &\\
		\Gamma\indices{^2_0_0}&=(C+O^ν)τe^{-2(p_\circ^1+p_\circ^3)\tau}\to 0\,, &
		\Gamma\indices{^2_1_0}&=(C+O^ν)\,, &\\
		\Gamma\indices{^2_1_1}&=(C+O^ν)τe^{2(-p_\circ^1+p_\circ^2)\tau}\,, &
		\Gamma\indices{^2_2_1}&=(C+O^ν)τ\,, &\\
		\Gamma\indices{^2_2_2}&=(C+O^ν)τ\,, &
		\Gamma\indices{^2_3_0}&=o(e^{2(p_\circ^2-p_\circ^3)\tau})\to 0\,, &\\
		\Gamma\indices{^2_3_1}&=(C+O^ν)τ\,, &
		\Gamma\indices{^2_3_2}&=(C+O^ν)τ\,, &\\
		\Gamma\indices{^2_3_3}&=(C+O^ν)τe^{2(p_\circ^2-p_\circ^3)\tau})\to 0\,, &
		\Gamma\indices{^3_0_0}&=(C+O^ν)τe^{2(-p_\circ^1+p_\circ^2)\tau}\to 0\,, &\\
		\Gamma\indices{^3_1_0}&=(C+O^ν)\,, &
		\Gamma\indices{^3_1_1}&=(C+O^ν)τe^{2(-p_\circ^1+p_\circ^3)\tau}\,, &\\
		\Gamma\indices{^3_2_0}&=(C+O^ν)\,, &
		\Gamma\indices{^3_2_1}&=(C+O^ν)τe^{2(-p_\circ^2+p_\circ^3)\tau}\,, &\\
		\Gamma\indices{^3_2_2}&=(C+O^ν)τe^{2(-p_\circ^2+p_\circ^3)\tau}\,, &
		\Gamma\indices{^3_3_1}&=(C+O^ν)τ\,, &\\
		\Gamma\indices{^3_3_2}&=(C+O^ν)τ\,, &
		\Gamma\indices{^3_3_3}&=(C+O^ν)τ\,,
		\end{align*}}
	with $C$ denoting $τ$-independent quantities, possibly different for different components. For most components an explicit first order term is given above. This is the case if the highest order term appearing in \eqref{2015VII6.1} does not vanish and therefore only the first order terms of the metric, given in \eqref{2015VII6.2} contribute. For the other terms, \eg $Γ^1{}_{20}$, the highest order term cancels and lower order terms of the metric could potentially become important, necessitating a more detailed analysis. Here the behaviour of these terms is simply given as decaying faster than the vanishing highest order term, as this is sufficient to determine the behaviour of the Kretschmann scalar in the following.
	
	From the Christoffels we calculate the behaviour of the coordinate components of the Riemann tensor, defined as
	\[
	R_{αβγδ}=\bar{g}_{ασ}\left(Γ^σ{}_{βδ,γ}-Γ^σ{}_{βγ,δ}+Γ^σ{}_{γε}Γ^ε{}_{βδ}-Γ^σ{}_{δε}Γ^ε{}_{βγ}\right)\,.
	\]
	Their behaviour is given by
	\begin{align*}
	R_{1010}&=(C +O^ν) e^{-\tau 2 p_\circ^1}\,,&
	R_{2010}&=(C +O^ν) e^{-\tau 2 p_\circ^2}\to0\,,&\\
	R_{2020}&=(C +O^ν) e^{-\tau 2 p_\circ^2}\to0\,,&
	R_{2110}&=(C +O^ν) \tau e^{-τ2p_\circ^1}\,,&\\
	R_{2120}&=(C +O^ν) \tau e^{-τ2p_\circ^2}\to0\,,&
	R_{2121}&=(C +O^ν) e^{τ2p_\circ^3}\,,&\\
	R_{3010}&=(C +O^ν) e^{-\tau 2 p_\circ^3}\to0\,,&
	R_{3020}&=(C +O^ν) e^{-τ 2 p_\circ^3}\to0\,,&\\
	R_{3021}&=o(\tau e^{-τ 2 p_\circ^2})\to0\,,&
	R_{3030}&=(C +O^ν) e^{-τ 2 p_\circ^3}\to0\,,&\\
	R_{3110}&=(C +O^ν) \tau e^{-\tau 2p_\circ^1}\,,&
	R_{3120}&=(C +O^ν) \tau e^{-\tau 2 p_\circ^2}\to0\,,&\\
	R_{3121}&=(C +O^ν) e^{\tau 2 p_\circ^2}\,,&
	R_{3130}&=(C +O^ν) \tau e^{-\tau 2 p_\circ^3}\to0\,,&\\
	R_{3131}&=(C +O^ν) e^{\tau 2 p_\circ^2}\,,&
	R_{3210}&=(C +O^ν) \tau e^{-\tau 2 p_\circ^2}\to0\,,&\\
	R_{3220}&=(C +O^ν) \tau e^{-\tau 2 p_\circ^2}\to0\,,&
	R_{3221}&=(C +O^ν) e^{\tau 2 p_\circ^1}\to0\,,&\\
	R_{3230}&=(C +O^ν) \tau e^{-\tau 2 p_\circ^3}\to0\,,&
	R_{3231}&=(C +O^ν) e^{\tau 2 p_\circ^1}\to0\,,&\\
	R_{3232}&=(C +O^ν) e^{\tau 2 p_\circ^1}\to0\,.&
	\end{align*}
	
	The Kretschmann scalar, defined as $R_{αβγδ}R^{αβγδ}$, behaves as
	\begin{equation}\label{kretsch}
	K=R_{αβγδ}R^{αβγδ}=(C^K+O(e^{-\nu\tau}))e^{\tau 4(p_\circ^1+p_\circ^2+p_\circ^3)}\,,
	\end{equation}
	with
	\begin{equation}\label{kretsch_koeff}
	C^K=\frac{16e^{4σ_{β_\circ}}}{(p_\circ^2+p_\circ^3)^2}
	\big(p_\circ^2 p_\circ^3\big)^2 \big((p_\circ^2)^2+p_\circ^2p_\circ^3+(p_\circ^3)^2\big)\,.
	\end{equation}
	As the inequalities \eqref{final_my_p_cond} require $p_\circ^3>p_\circ^2>0$ the coefficient $C^K$ is positive and the Kretschmann scalar diverges as $τ\to\infty$ for all constructed solutions.
	
	\section{Solutions including a cosmological constant}\label{sec:cosm_const}
	Solutions of the form presented above can also be constructed with a nonzero cosmological constant. Here we will show the resulting changes in the arguments above.
	
	The presence of a cosmological constant changes the action to
	\begin{equation}\label{2015VII21.1}
		S[\bar{g}_{\mu\nu}]=\int\D^Dx \sqrt{-\bar{g}}(\bar{R}-2\Lambda)\,.
	\end{equation}
	This leads to an additional term $2\Lambda g$ in the Hamiltonian. As the determinant $g$ of the spatial metric behaves asymptotically as
	\begin{equation}
	g=e^{-2(\sigma_{p_\circ}\tau+\sigma_{\beta_\circ})}(1+O(e^{-\nu\tau}))
	\end{equation}
	the term decays exponentially. It contributes an additional term to the evolution equation for $\bar{\pi}_a$, \eqref{my_pi_eq}. This equation includes the diverging prefactor $e^{\epsilon \tau}$ but the conditions \eqref{my_conds} guarantee that the new term decays fast enough to compensate it. Therefore the presence of a cosmological constant introduces no new conditions on the free functions from the evolution equations.
	
	The cosmological constant also appears in the Hamiltonian constraint, but not in the momentum constraint. As the additional term in the Hamiltonian decays it does not change the asymptotic Hamiltonian constraint or the condition on the $p_\circ^a$ arising from it.
	
	Finally, $\Lambda$ appears in the derivation of the evolution equations for the constraints in \ref{sec:const_evo}. Equations \eqref{17III15.8} and \eqref{17III15.10} change respectively to
	\begin{align}
		H&=-\frac{2}{\tilde{N}^2}G_{00}+2g\Lambda+O(N^k)\,,\\
		\bar{G}^{ij}&=-\frac{1}{2g}Hg^{ij}-\Lambda g^{ij}+O(N^k)\,.
	\end{align}
	The additional terms containing $\Lambda$ cancel in the final evolution equations \eqref{17III15.11} and \eqref{18III15.1}. Therefore the arguments regarding the constraints in section \ref{sec:asym_full_const} remain unchanged.
	
	We conclude that the solutions with the asymptotic behaviour  given in \ref{sec:asym_beh}, as obtained from the Fuchs theorem above, exist for all values of $\Lambda$, with the free functions and the associated conditions unchanged from the $\Lambda=0$ case.
	
	\newpage
	\bibliographystyle{elsarticle-num.bst}
	\bibliography{bib.bib}

\begin{thebibliography}{10}
\expandafter\ifx\csname url\endcsname\relax
  \def\url#1{\texttt{#1}}\fi
\expandafter\ifx\csname urlprefix\endcsname\relax\def\urlprefix{URL }\fi
\expandafter\ifx\csname href\endcsname\relax
  \def\href#1#2{#2} \def\path#1{#1}\fi

\bibitem{Eisenstaedt1989}
J.~Eisenstaedt, The early interpretation of the schwarzschild solution, in:
  D.~Howard, J.~Stachel (Eds.), Einstein and the History of General Relativity,
  Birkh\"a{}user, 1989, pp. 213--234.

\bibitem{Painleve1921}
P.~{Painlevé}, {La mécanique classique et la théorie de la relativité}, C.
  R. Acad. Sci. (Paris) 173 (1921) 677--680.

\bibitem{Gullstrand1922}
A.~{Gullstrand}, {Allgemeine Lösung des statischen Einkörperproblems in der
  Einsteinschen Gravitationstheorie}, Arkiv. Mat. Astron. Fys. 16(8) (1922)
  1--15.

\bibitem{Lemaitre1932}
G.~{Lema{\^i}tre}, {L'Univers en expansion}, Annales de la Soci{\'e}t{\'e}
  Scientifique de Bruxelles 53 (1933) 51.

\bibitem{Penrose1969}
R.~{Penrose}, {Gravitational Collapse: the Role of General Relativity}, Nuovo
  Cimento Rivista Serie 1 (1969) 252.
\newblock \href {http://dx.doi.org/10.1023/A:1016578408204}
  {\path{doi:10.1023/A:1016578408204}}.

\bibitem{Penrose1965}
R.~Penrose, Gravitational collapse and space-time singularities, Phys. Rev.
  Lett. 14 (1965) 57--59.
\newblock \href {http://dx.doi.org/10.1103/PhysRevLett.14.57}
  {\path{doi:10.1103/PhysRevLett.14.57}}.

\bibitem{Hawking1970}
S.~W. Hawking, R.~Penrose, The singularities of gravitational collapse and
  cosmology, Proceedings of the Royal Society of London A: Mathematical,
  Physical and Engineering Sciences 314~(1519) (1970) 529--548.
\newblock \href {http://dx.doi.org/10.1098/rspa.1970.0021}
  {\path{doi:10.1098/rspa.1970.0021}}.

\bibitem{Lifschitz1963}
E.~Lifschitz, I.~Khalatnikov, Investigations in relativistic cosmology,
  Advances in Physics 12~(46) (1963) 185--249.
\newblock \href {http://dx.doi.org/10.1080/00018736300101283}
  {\path{doi:10.1080/00018736300101283}}.

\bibitem{Belinski1970}
V.~Belinski, I.~Khalatnikov, E.~Lifschitz, {Oscillatory approach to a singular
  point in the relativistic cosmology}, Advances in Physics 19 (1970) 525--573.
\newblock \href {http://dx.doi.org/10.1080/00018737000101171}
  {\path{doi:10.1080/00018737000101171}}.

\bibitem{Chitre1972}
D.~M. Chitre, Investigation of vanishing of a horizon for bianchi type ix (the
  mixmaster) universe, Ph.D. thesis, University of Maryland (1972).

\bibitem{Misner1994}
C.~W. Misner, {The Mixmaster cosmological metrics}, D. Hobill, ed.,
  Deterministic chaos in general relativity, Plenum (1994) 317--328\href
  {http://arxiv.org/abs/gr-qc/9405068} {\path{arXiv:gr-qc/9405068}}.

\bibitem{Damour2003}
T.~Damour, M.~Henneaux, H.~Nicolai, {Cosmological billiards}, Classical and
  Quantum Gravity\href {http://arxiv.org/abs/hep-th/0212256}
  {\path{arXiv:hep-th/0212256}}, \href
  {http://dx.doi.org/10.1088/0264-9381/20/9/201}
  {\path{doi:10.1088/0264-9381/20/9/201}}.

\bibitem{Berger:2000uf}
B.~K. Berger, V.~Moncrief, {Exact $U(1)$ symmetric cosmologies with local
  mixmaster dynamics}, Phys.Rev. D62 (2000) 023509.
\newblock \href {http://arxiv.org/abs/gr-qc/0001083}
  {\path{arXiv:gr-qc/0001083}}, \href
  {http://dx.doi.org/10.1103/PhysRevD.62.023509}
  {\path{doi:10.1103/PhysRevD.62.023509}}.

\bibitem{Garfinkle2004}
D.~{Garfinkle}, {Numerical Simulations of Generic Singularities}, Physical
  Review Letters 93~(16) (2004) 161101.
\newblock \href {http://arxiv.org/abs/gr-qc/0312117}
  {\path{arXiv:gr-qc/0312117}}, \href
  {http://dx.doi.org/10.1103/PhysRevLett.93.161101}
  {\path{doi:10.1103/PhysRevLett.93.161101}}.

\bibitem{Garfinkle2007}
D.~Garfinkle, {Numerical simulations of general gravitational singularities},
  Class.Quant.Grav. 24 (2007) S295--S306.
\newblock \href {http://arxiv.org/abs/0808.0160} {\path{arXiv:0808.0160}},
  \href {http://dx.doi.org/10.1088/0264-9381/24/12/S19}
  {\path{doi:10.1088/0264-9381/24/12/S19}}.

\bibitem{Rendall2001}
A.~D. Rendall, M.~Weaver, {Manufacture of Gowdy space-times with spikes},
  Class.Quant.Grav. 18 (2001) 2959--2976.
\newblock \href {http://arxiv.org/abs/gr-qc/0103102}
  {\path{arXiv:gr-qc/0103102}}, \href
  {http://dx.doi.org/10.1088/0264-9381/18/15/310}
  {\path{doi:10.1088/0264-9381/18/15/310}}.

\bibitem{Belinski1973}
V.~Belinski, I.~Khalatnikov, {Effect of Scalar and Vector Fields on the Nature
  of the Cosmological Singularity}, Sov.Phys.JETP 36 (1973) 591.

\bibitem{Andersson2001}
L.~Andersson, A.~Rendall, {Quiescent cosmological singularities},
  Communications in Mathematical Physics (2001) 1--38\href
  {http://arxiv.org/abs/gr-qc/0001047} {\path{arXiv:gr-qc/0001047}}, \href
  {http://dx.doi.org/10.1007/s002200100406} {\path{doi:10.1007/s002200100406}}.

\bibitem{Damour2002}
T.~Damour, M.~Henneaux, A.~Rendall, M.~Weaver, {Kasner-like behaviour for
  subcritical Einstein-matter systems}, Annales Henri Poincar\'{e}\href
  {http://arxiv.org/abs/gr-qc/0202069v2} {\path{arXiv:gr-qc/0202069v2}}, \href
  {http://dx.doi.org/10.1007/s000230200000} {\path{doi:10.1007/s000230200000}}.

\bibitem{Demaret1985}
J.~Demaret, M.~Henneaux, P.~Spindel, {Non-oscillatory behaviour in vacuum
  Kaluza-Klein cosmologies}, Physics Letters B 164~(December) (1985) 27--30.
\newblock \href {http://dx.doi.org/10.1016/0370-2693(85)90024-3}
  {\path{doi:10.1016/0370-2693(85)90024-3}}.

\bibitem{Isenberg:1989gq}
J.~Isenberg, V.~Moncrief, {Asymptotic behavior of the gravitational field and
  the nature of singularities in Gowdy space-times}, Annals Phys. 199 (1990)
  84--122.
\newblock \href {http://dx.doi.org/10.1016/0003-4916(90)90369-Y}
  {\path{doi:10.1016/0003-4916(90)90369-Y}}.

\bibitem{Chrusciel:1990wn}
P.~Chru\'sciel, J.~Isenberg, V.~Moncrief, {Strong cosmic censorship in
  polarized Gowdy space-times}, Class.Quant.Grav. 7 (1990) 1671--1680.
\newblock \href {http://dx.doi.org/10.1088/0264-9381/7/10/003}
  {\path{doi:10.1088/0264-9381/7/10/003}}.

\bibitem{Kichenassamy1998}
S.~Kichenassamy, A.~Rendall, {Analytic description of singularities in Gowdy
  spacetimes}, Classical and Quantum Gravity 1339.
\newblock \href {http://dx.doi.org/10.1088/0264-9381/15/5/016}
  {\path{doi:10.1088/0264-9381/15/5/016}}.

\bibitem{Isenberg1999}
J.~Isenberg, S.~Kichenassamy, {Asymptotic behavior in polarized $T^2$-symmetric
  vacuum space–times}, Journal of Mathematical Physics 40~(1) (1999) 340.
\newblock \href {http://dx.doi.org/10.1063/1.532775}
  {\path{doi:10.1063/1.532775}}.

\bibitem{Clausen2007}
A.~Clausen, J.~Isenberg, {Areal Foliation and AVTD Behavior in $T^2$ Symmetric
  Spacetimes with Positive Cosmological Constant}, J. Math. Phys (2007)
  1--15\href {http://arxiv.org/abs/gr-qc/0701054v2}
  {\path{arXiv:gr-qc/0701054v2}}, \href {http://dx.doi.org/10.1063/1.2767534}
  {\path{doi:10.1063/1.2767534}}.

\bibitem{Ames2013}
E.~Ames, F.~Beyer, J.~Isenberg, P.~LeFloch, {Quasilinear Hyperbolic Fuchsian
  Systems and AVTD Behavior in $T^2$-Symmetric Vacuum Spacetimes}, Annales
  Henri Poincar\'{e}~(Paris 6) (2013) 1--78.
\newblock \href {http://arxiv.org/abs/1205.1881v2} {\path{arXiv:1205.1881v2}},
  \href {http://dx.doi.org/10.1007/s00023-012-0228-2}
  {\path{doi:10.1007/s00023-012-0228-2}}.

\bibitem{Isenberg2002}
J.~Isenberg, V.~Moncrief, {Asymptotic behaviour in polarized and half-polarized
  $U(1)$ symmetric vacuum spacetimes}, Classical and Quantum Gravity~(1).
\newblock \href {http://arxiv.org/abs/gr-qc/0203042}
  {\path{arXiv:gr-qc/0203042}}, \href
  {http://dx.doi.org/10.1088/0264-9381/19/21/305}
  {\path{doi:10.1088/0264-9381/19/21/305}}.

\bibitem{ChoquetBruhat:2005xs}
Y.~Choquet-Bruhat, J.~Isenberg, V.~Moncrief, {Topologically general $U(1)$
  symmetric Einstein spacetimes with AVTD behavior}, Nuovo Cim. B119 (2004)
  625--638.
\newblock \href {http://arxiv.org/abs/gr-qc/0502104}
  {\path{arXiv:gr-qc/0502104}}, \href
  {http://dx.doi.org/10.1393/ncb/i2004-10174-x}
  {\path{doi:10.1393/ncb/i2004-10174-x}}.

\bibitem{ChoquetBruhat:2005rd}
Y.~Choquet-Bruhat, J.~Isenberg, {Half polarized $U(1)$ symmetric vacuum
  spacetimes with AVTD behavior}, J.Geom.Phys. 56 (2006) 1199--1214.
\newblock \href {http://arxiv.org/abs/gr-qc/0506066}
  {\path{arXiv:gr-qc/0506066}}, \href
  {http://dx.doi.org/10.1016/j.geomphys.2005.06.011}
  {\path{doi:10.1016/j.geomphys.2005.06.011}}.

\bibitem{Ringstrom2008}
H.~Ringström, Strong cosmic censorship in the case of $t^3$ -gowdy vacuum
  spacetimes, Classical and Quantum Gravity 25~(11).
\newblock \href {http://dx.doi.org/10.1088/0264-9381/25/11/114010}
  {\path{doi:10.1088/0264-9381/25/11/114010}}.

\bibitem{Damour2008}
T.~Damour, S.~de~Buyl, {Describing general cosmological singularities in
  Iwasawa variables}, Physical Review D\href {http://arxiv.org/abs/0710.5692}
  {\path{arXiv:0710.5692}}, \href
  {http://dx.doi.org/10.1103/PhysRevD.77.043520}
  {\path{doi:10.1103/PhysRevD.77.043520}}.

\bibitem{Wald1984}
R.~Wald, General Relativity, University of Chicago Press, 1984.

\bibitem{Choquet-Bruhat2008}
Y.~Choquet-Bruhat, {General relativity and the Einstein equations}, Oxford
  University Press, 2008.

\end{thebibliography}
\end{document}